\documentclass[10pt,twocolumn,twoside]{IEEEtran}
\usepackage{ifpdf}
\usepackage{url}
\ifpdf
\else
\fi
\usepackage{pifont}
\usepackage{cite}
\usepackage{balance}
\usepackage{bm,comment,color}

\ifCLASSINFOpdf
  \usepackage[pdftex]{graphicx}
  \graphicspath{{../pdf/}{../jpeg/}}
  \DeclareGraphicsExtensions{.pdf,.jpeg,.png}
\else
  \usepackage[dvips]{graphicx}
  \graphicspath{{../eps/}}
  \DeclareGraphicsExtensions{.eps}
\fi
\usepackage{amsmath}
\usepackage{algpseudocode}
\algtext*{EndWhile}
\algtext*{EndIf}
\algtext*{EndFor}
\usepackage{algorithm}
\usepackage{array}
\usepackage{makecell}
\usepackage{amsfonts} 
\usepackage{amssymb}
\usepackage{esint} 
\usepackage{units}
\usepackage{multirow}
\usepackage{comment}

\usepackage{amsthm}

\newtheorem{Prob}{Problem}

\usepackage{color}

\usepackage{standalone}

\usepackage{pgfplots}
\usepackage{tikz}
\usetikzlibrary{decorations.pathreplacing}
\usetikzlibrary{calc}
\makeatletter
\newcommand{\gettikzxy}[3]{%
  \tikz@scan@one@point\pgfutil@firstofone#1\relax
  \edef#2{\the\pgf@x}%
  \edef#3{\the\pgf@y}%
}
\usetikzlibrary{spy,backgrounds}
\pgfplotsset{compat=newest}
\usetikzlibrary{plotmarks}
\usetikzlibrary{arrows.meta}
\usepgfplotslibrary{patchplots}
\usepackage{grffile}
\newlength\fheight 
\newlength\fwidth 
\usepgfplotslibrary{fillbetween}

\usepackage{acronym}

\acrodef{6g}[6G]{the sixth generation}
\acrodef{ae}[AE]{autoencoder}
\acrodef{lqr}[LQR]{linear quadratic regulator}
\acrodef{aoa}[AOA]{angle-of-arrival}
\acrodef{bs}[BS]{base station}
\acrodef{bse}[BSE]{beam squint effect}
\acrodef{crb}[CRB]{Cram\'er-Rao bound}
\acrodef{dt}[DT]{digital twin}
\acrodef{elaa}[ELAA]{extremely large antenna array}
\acrodef{ff}[FF]{far-field}
\acrodef{gru}[GRU]{gated recurrent unit}
\acrodef{isac}[ISAC]{integrated sensing and communication}
\acrodef{las}[L\&S]{localization and sensing}
\acrodef{los}[LOS]{line-of-sight}
\acrodef{nf}[NF]{near-field}
\acrodef{nlos}[NLOS]{non-line-of-sight}
\acrodef{ofdm}[OFDM]{orthogonal frequency division multiplexing}
\acrodef{ris}[RIS]{reconfigurable intelligent surface}
\acrodef{scc}[SCC]{sensing–communication–control}
\acrodef{sns}[SNS]{spatial non-stationarity}
\acrodef{swm}[SWM]{spherical wave model}
\acrodef{siso}[SISO]{single-input single-output}
\acrodef{ue}[UE]{user equipment}
\acrodef{dmimo}[D-MIMO]{distributed MIMO}
\acrodef{ppo}[PPO]{proximal policy optimization}

\usepackage{tabu,longtable}

\ifCLASSOPTIONcompsoc
 \usepackage[caption=false,font=normalsize,labelfont=sf,textfont=sf]{subfig}
\else
 \usepackage[caption=false,font=footnotesize]{subfig}
\fi
\usepackage{stfloats}
\usepackage[hidelinks]{hyperref}
\usepackage{xcolor}
\long\def\comment#1{}

\newfont{\bbb}{msbm10 scaled 700}

\newfont{\bb}{msbm10 scaled 1100}

\definecolor{darkred}{RGB}{180,0,0} 
\usepackage{graphicx}      

\allowdisplaybreaks

\renewcommand{\baselinestretch}{0.99} 
\usepackage{booktabs}

\usepackage{amsthm}

\usepackage{titlesec}
\titlespacing{\subsection}{0pt}{*0.5}{*0.3}
\titlespacing{\section}{0pt}{*0.5}{*0.3}

\renewcommand{\thesubsubsection}{\arabic{subsubsection})}
\makeatletter
\@addtoreset{subsubsection}{subsection}
\makeatother
\titleformat{\subsubsection}[runin]
  {\normalfont\normalsize\itshape}      %
  {\quad \thesubsubsection}          %
  {0.5em}                       %
  {}[:\hspace{0.5em}]                         %
\titlespacing{\subsubsection}{0pt}{*0.5}{1em} %

\usepackage{caption}
\usepackage{subcaption}
\begin{document} 
\bstctlcite{IEEEexample:BSTcontrol}

\title{ Semantic Communication for Rate-Limited Closed-Loop Distributed Communication-Sensing-Control Systems}

\author{
Guangjin Pan, \IEEEmembership{Member, IEEE}, Ayça Özçelikkale, \IEEEmembership{Member, IEEE},
Christian Häger, \IEEEmembership{Member, IEEE}, \\ Musa Furkan Keskin, \IEEEmembership{Member, IEEE}, Henk Wymeersch, \IEEEmembership{Fellow, IEEE}
\thanks{This work was supported in part by a grant from the Chalmers AI Research
 Center Consortium (CHAIR), by the National Academic Infrastructure for
 Supercomputing in Sweden (NAISS), by the SNS JU project 6G-DISAC under the EU's Horizon Europe research and innovation Program under Grant Agreement No 101139130,  the Swedish Foundation for Strategic Research (SSF) (grant FUS21-0004, SAICOM), the Swedish Research Council (VR) through the project 6G-PERCEF under Grant 2024-04390, and the Chalmers Areas of Advance in ICT and Transport. (Corresponding author: Guangjin Pan)}
\thanks{Guangjin Pan, Christian Häger, Musa Furkan Keskin, and Henk Wymeersch are with Department of Electrical Engineering, Chalmers University of Technology, 41296 Gothenburg, Sweden (e-mail: {guangjin.pan; christian.haeger; furkan; henkw}@chalmers.se).}
\thanks{Ayça Özçelikkale is with Department of Electrical Engineering, Uppsala University, Uppsala, Sweden (e-mail: ayca.ozcelikkale@angstrom.uu.se).}}

\maketitle

\begin{abstract}
The growing integration of distributed integrated sensing and communication (ISAC) with closed-loop control in intelligent networks demands efficient information transmission under stringent bandwidth constraints.
To address this challenge, this paper proposes a unified framework for goal-oriented semantic communication in distributed \ac{scc} systems.
Building upon Weaver’s three-level model, we establish a hierarchical semantic formulation with three error levels (L1: observation reconstruction, L2: state estimation, and L3: control) to jointly optimize their corresponding objectives. Based on this formulation, we propose a unified goal-oriented semantic compression and rate adaptation framework that is applicable to different semantic error levels and optimization goals across the \ac{scc} loop. A rate-limited multi-sensor \ac{lqr} system is used as a case study to validate the proposed framework. We employ a \ac{gru}–based \ac{ae} for semantic compression and a \ac{ppo}–based rate adaptation algorithm that dynamically allocates transmission rates across sensors.
{Results show that the proposed framework effectively captures task-relevant semantics and adapts its resource allocation strategies across different semantic levels, thereby achieving level-specific performance gains under bandwidth constraints.}
\end{abstract}


\IEEEpeerreviewmaketitle
\acresetall 
\section{Introduction}

\begin{figure}[tb]
\centering
 \includegraphics[scale=0.32]{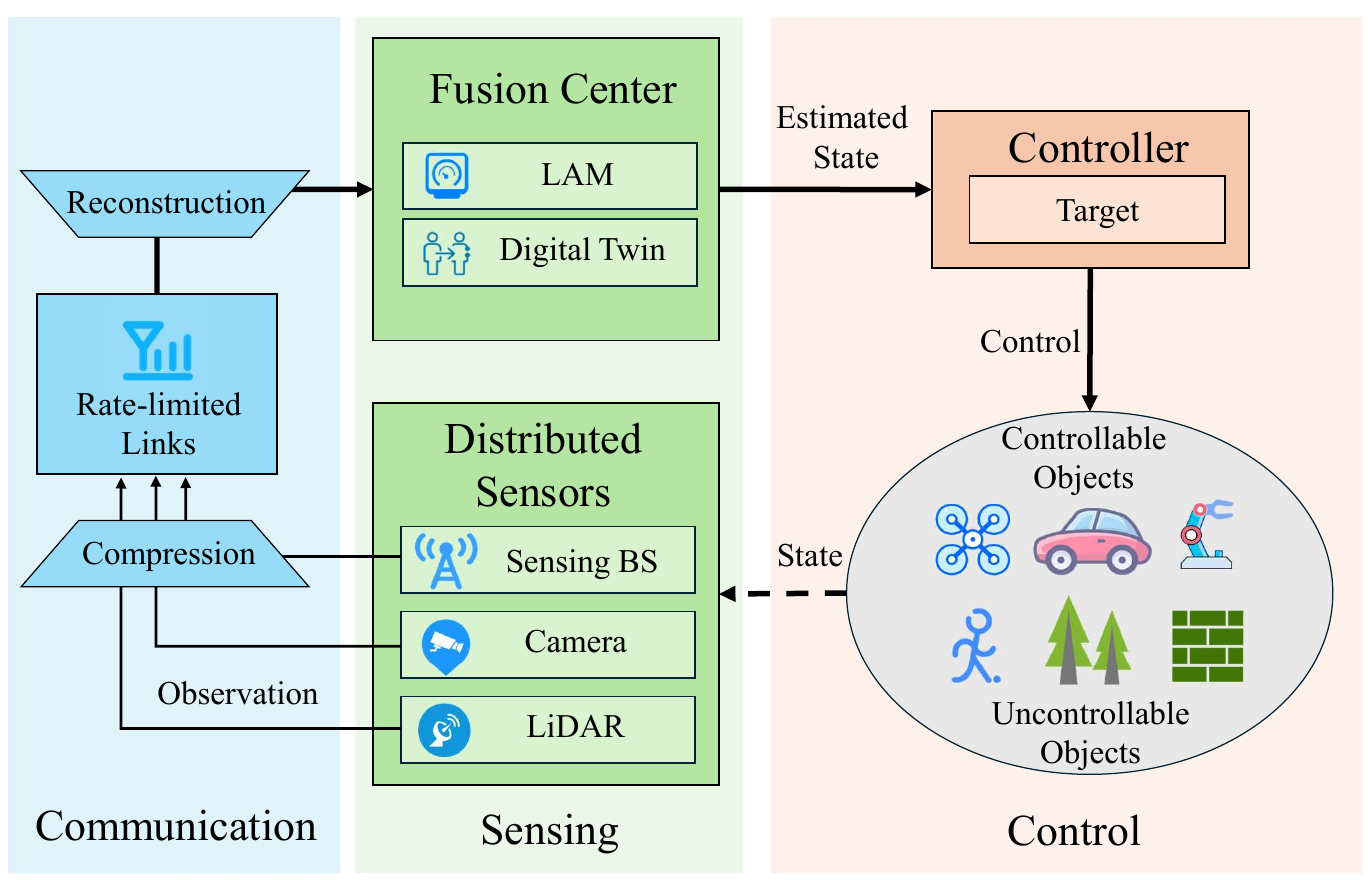}
\caption{System architecture of a closed-loop distributed \ac{scc} system with rate-limited links.}
\label{fig:DISACsystem}
\end{figure}

\Ac{isac} has emerged as a key enabler for next-generation wireless networks, 
where a shared infrastructure jointly supports high-resolution sensing and reliable communication~\cite{Strinati_DISAC_2025}. 
While many existing \ac{isac} studies focus on integrating sensing and communication functionalities, future applications such as autonomous driving, low-altitude economy, and industrial automation require the joint design of sensing, communication, and control to enable truly closed-loop intelligence~\cite{Park_WNCS_2018, Chang_IntegratedUAV_2022, pan2025active}.
This trend motivates the development of closed-loop \ac{isac} systems, which inherently embody the concept of \ac{scc} co-design, a holistic paradigm that explicitly captures the interdependence among these three subsystems~\cite{Park_WNCS_2018}.
In such closed-loop systems (as illustrated in Fig.~\ref{fig:DISACsystem}), sensors continuously observe the environment and transmit their measurements to a fusion center for state estimation.
Based on the estimated states, the controllers then compute appropriate control actions and apply them to the corresponding targets.
This closed-loop integration is crucial for latency-sensitive and safety-critical applications, as control performance depends on sensing accuracy and communication reliability~\cite{cao2024goal}.

In wireless control systems~\cite{ballotta2025role, park2017wireless,branz2020time}, the wireless links are often bandwidth-limited, which may result in unacceptable transmission delays or packet losses, severely degrading the closed-loop performance~\cite{Park_WNCS_2018}. This problem becomes particularly critical when sensors transmit their observations to the fusion center for state estimation.
Since the observations are typically high-dimensional (such as wireless channel measurements~\cite{pan2025ai}, camera images~\cite{zhang2023predictive}, or dense point clouds\cite{shao2025point}), the required data rate may exceed the transmission capacity of the wireless network.
The challenge is further exacerbated in systems with multiple distributed sensors, where concurrent transmissions can easily cause network congestion, additional latency, and information loss~\cite{cao2024goal_mag}.

To address these challenges, early studies focus on networked control systems (NCSs)~\cite{ballotta2025role, park2017wireless,branz2020time}, where communication imperfections such as packet loss~\cite{cetinkaya2016networked, li2023dynamic, zhu2023output} and delay~\cite{yan2017networked, zhu2023output, ballotta2025role} are explicitly incorporated into controller design.
To improve communication efficiency under limited bandwidth, subsequent research introduces event-triggered and resource-scheduling schemes~\cite{li2023dynamic, zhu2023output, ren2022event}, where control updates or sensor transmissions are triggered only when significant state deviations occur.
For example,~\cite{li2023dynamic} presents a co-design method that combines a dynamic periodic event-triggering mechanism with an output-based control law for NCSs subject to successive packet dropouts. 
In addition, several works~\cite{li2021distributed, shibata2021deep} employ reinforcement learning (RL)–based scheduling to alleviate the challenges of network congestion in dynamic communication environments. However, such methods usually rely on discarding part of the information to meet bandwidth constraints, which inevitably leads to degraded state estimation and control performance.

Another possible solution relies on information compression.
Sensors can compress their observation before transmission to meet bandwidth and latency constraints.
Typical compression techniques include principal component analysis (PCA) and \ac{ae}–based feature extraction~\cite{pan2025rate, Nagashima_PCA_2016, Ismayilov_AE_2024}, which reduce dimensionality via linear projection onto the principal subspace or nonlinear latent coding optimized under a reconstruction objective. 
However, these methods are primarily reconstruction-oriented, focusing on compressing observations and minimizing reconstruction distortion.
Such an approach may waste bandwidth by transmitting information that is irrelevant to downstream estimation or control tasks.
Recently, semantic communication has attracted increasing attention due to its potential to enable task-driven and communication-efficient information exchange~\cite{Yang2023Semantic,sagduyu2024will, gunduz2022guest}.
Its superiority has been extensively demonstrated in various application domains such as text, image, and video transmission.
For instance, the authors in~\cite{xie2021deep} design a deep learning–based semantic communication system for text transmission.
For visual data, some works~\cite{zhang2023predictive, jiang2022wireless} propose semantic encoders for image and video transmission, showing that semantic feature extraction can achieve robust performance.

Building upon these advances, goal-oriented semantic communication has been proposed to further optimize the communication process according to the ultimate task objective rather than intermediate reconstruction fidelity. It has been applied to \ac{scc} systems~\cite{wu2024goal,li2025integrated,cao2024goal,girgis2023semantic,li2025unified,fang2025sensing}. For instance, in~\cite{wu2024goal}, the authors propose a goal-oriented semantic communication framework for robotic control, which exploits the contextual relevance and importance of data at both the transmitter and receiver to enhance task performance.
Similarly, the authors in~\cite{li2025integrated} introduce a sequences-based semantic compression scheme for control information, enabling adaptive robot control.
However,~\cite{wu2024goal, li2025integrated} do not consider closed-loop control systems.
Considering a closed-loop \ac{scc} system, the authors in~\cite{cao2024goal} propose a goal-oriented co-design framework for wireless NCSs with bidirectional links, where a binary transmission policy jointly optimizes estimation, scheduling, and control to minimize state violation probability under resource constraints. The framework in~\cite{girgis2023semantic} also considers a closed-loop control setting, where semantic representation and logical control decisions are jointly optimized within the feedback loop of correlated dynamical systems. The authors in~\cite{li2025unified} develop a unified timing analysis framework for closed-loop goal-oriented communication, providing insights into how stochastic transmission delays and update intervals influence control performance.  In~\cite{fang2025sensing}, the authors extend this concept to sensing–communication–computing–control systems and propose a goal-oriented closed-loop optimization framework that jointly allocates bandwidth and computing resources to minimize the total \ac{lqr} cost. {However, despite these remarkable contributions to goal-oriented semantic communication for closed-loop systems,
the works in~\cite{li2025unified, fang2025sensing} focus primarily on theoretical timing and delay analyses,
while~\cite{cao2024goal, girgis2023semantic} design goal-oriented transmission schemes oriented toward state or control signals.
None of these studies explicitly address the communication burden arising from transmitting high-dimensional sensing observations, which often dominate bandwidth consumption in distributed \ac{scc} systems. In contrast, our work investigates goal-oriented semantic communication from the perspective of sensor observations,
analyzing how semantic compression and rate adaptation jointly affect observation reconstruction, state estimation, and control performance. This perspective provides new insights into achieving efficient closed-loop coordination under communication constraints.}

In closed-loop \ac{scc} systems, the meaning of transmitted information from the sensors is inherently task-dependent. It depends on how communication affects state estimation and control decisions, rather than on how accurately the raw signals are reconstructed.
This consideration becomes even more critical in distributed \ac{scc} networks, where multiple sensors observe different aspects of the environment under limited communication resources. In such scenarios, it is essential not only to extract task-relevant semantic information from each sensor’s observation, but also to balance the semantic importance across sensors, so that the network prioritizes the transmission of more critical information within constrained bandwidth.
These challenges motivate the development of a unified closed-loop \ac{scc} framework that models the system under different goals and provides a general solution approach for semantic compression and rate allocation among distributed sensors. Specifically, the main contributions of this paper are summarized as follows:
\begin{itemize}
    \item \textbf{Unified hierarchical semantic formulation for closed-loop distributed \ac{scc} systems:} 
 We develop a unified modeling framework for closed-loop distributed \ac{scc} systems that explicitly captures the interdependence among sensing, communication, and control. 
    Building upon Weaver’s three-level communication model, we establish a hierarchical semantic formulation that bridges semantic communication with closed-loop control theory. 
    Three semantic error levels are defined: L1 (observation reconstruction), L2 (state estimation), and L3 (control), corresponding respectively to minimizing observation distortion, state estimation error, and control cost.

\item \textbf{Unified goal-oriented semantic compression and rate adaptation framework:}  
    Based on the proposed semantic formulation, we develop a unified goal-oriented framework applicable to different semantic levels and task objectives across the \ac{scc} loop.  
    The framework integrates two key modules: a \emph{semantic compression module} that learns task-dependent latent representations across different semantic levels, and a \emph{rate adaptation module} that dynamically adjusts per-sensor transmission rates according to system states, communication conditions, and semantic objectives to satisfy the bandwidth constraint.

\item \textbf{LQR-based implementation:} 
Using a rate-limited \ac{lqr} system as a case study,  
we implement and evaluate the proposed framework to demonstrate its effectiveness in closed-loop \ac{scc} systems. Based on the proposed goal-oriented semantic communication framework, we develop a \ac{gru}-based \ac{ae} architecture for semantic compression  
and design a \ac{ppo}-based rate adaptation algorithm.  
Both modules are trained under the proposed three-level semantic error framework,  
with each level focusing respectively on optimizing observation reconstruction, estimation, and control objectives.
Experimental results verify that the hierarchical semantic representation effectively captures task-relevant information across different optimization goals. 
In particular, for optimizing closed-loop \ac{scc} system performance, modeling the L3 control level is essential, as it directly aligns communication with control objectives. 
Furthermore, the proposed GRU-AE+PPO algorithm significantly reduces the \ac{lqr} cost, achieving negligible performance loss compared with the rate-unconstrained baseline at L3. 
Meanwhile, the learned rate allocation strategies reveal distinct behaviors across semantic levels.
It is worth noting that although the algorithms are demonstrated on an \ac{lqr} system, the proposed framework and methods are general and can be extended to more complex nonlinear and multi-agent \ac{scc} systems.
\end{itemize}

The remainder of this paper is organized as follows.
Sec. ~II presents the closed-loop distributed SCC system model.
Building upon this foundation, Sec.~III introduces the unified hierarchical semantic formulation and proposes the unified goal-oriented semantic compression and rate adaptation framework.
In Sec.~IV, a rate-limited LQR system is employed as a case study, where GRU-AE–based semantic compression and PPO–based rate adaptation algorithms are developed within the proposed framework.
Sec.~V provides experimental results and performance analysis, and Sec.~VI concludes the paper.

\section{Closed-loop Distributed \ac{scc} System Model}

We consider a closed-loop distributed \ac{scc} system, as illustrated in Fig.~\ref{fig:DISACsystem}, where sensing, communication, and control are tightly coupled to enable real-time operation.  
Following a discrete-time approach, at each time step, the physical process evolves to a new state, which is 
observed by distributed sensors. The high-dimensional sensor data are locally compressed before being transmitted over rate-limited links to a fusion center. The fusion center reconstructs the data and produces a state estimate based on a \ac{dt} of the environment. Based on these estimates, the controller generates control actions that are fed back to the physical process, thereby closing the loop. In the following, we introduce a general probabilistic model of the \ac{scc} system, which serves as the foundation for subsequent algorithm design.



\subsection{State Model}
We consider a dynamic environment that contains both controllable objects and uncontrollable objects. 
The controllable objects refer to targets that are operated through remote control from a central server (e.g., vehicles), 
whereas the uncontrollable objects cannot be directly controlled by the server and include both static obstacles and autonomously moving objects (e.g., pedestrians). Accordingly, the overall system state at time $t$ can be decomposed as $\bm{x}_t =
[
(\bm{x}^{\text{ctl}}_t)^\top, 
(\bm{x}^{\text{unctl}}_t)^\top
]^\top$, 
where $\bm{x}^{\text{ctl}}_t$ denotes the state of controllable objects and $\bm{x}^{\text{unctl}}_t$ denotes the state of uncontrollable objects. Given control command $\bm{u}_{t}$, the joint system state at time $t$ is denoted by $\bm{x}_t$, which evolves according to a first-order Markov process:
$\bm{x}_{t+1}\sim p_{\text{obj}}(\bm{x}_{t+1}|\bm{x}_{t},\bm{u}_{t})$, 
where $p_{\text{obj}}(\cdot)$
represents the object dynamics and response to the control command $\bm{u}_{t}$. Note that the overall control input $\bm{u}_t$ also consists of two components,
$\bm{u}_t = \big[\bm{u}_t^{\text{ctl}}, \, \bm{u}_t^{\text{unctl}}\big]$,
where $\bm{u}_t^{\text{ctl}}$ denotes the control actions generated by the controller for the controllable objects, while $\bm{u}_t^{\text{unctl}}$ represents the autonomous or self-driven behaviors of uncontrollable objects (which may require no control input for static objects).

\subsection{Sensor Model}
The environment is equipped with $S$ heterogeneous sensors (e.g., sensing base stations (BSs), cameras, LiDARs), which continuously monitor and both controllable and uncontrollable objects. Each sensor provides high-dimensional raw observations, such as channel measurements, images, or dense point clouds. These measurements are transmitted to a fusion center for state estimation and prediction. 
Formally, the observation from sensor $s \in \{1,\ldots,S\}$ at time $t$ is modeled as
$
\bm{y}_{t}^{s} \sim p_{\text{meas}}^{s}\!\left(\bm{y}_{t}^{s}\mid \bm{x}_{t}\right)$,
where $p_{\text{meas}}^{s}(\cdot \mid \bm{x}_{t})$ denotes the sensor-specific observation model conditioned on the system state $\bm{x}_t$.

\subsection{Rate-limited Link}
In distributed \ac{scc} systems, communication links are often bandwidth-constrained, making the direct transmission of raw sensor observations $\bm{y}_{t}^{1:S}$ infeasible~\cite{Strinati_DISAC_2025, Guo_Toward_2025, pan2025ai}. Instead of forwarding all raw data, each sensor compresses its high-dimensional observation into a compact representation $\bm{z}_t^s$, subject to a long-term communication constraint $\frac{1}{T}\sum_{t=1}^T \sum_{s\in \mathcal{S}} R_t^s \le R$, where $R_t^s$ denotes the number of bits used to transmit $\bm{z}_t^s$, and $R$ represents the average total bandwidth budget over the time horizon~$T$. 
The compression process is modeled as a compression function: $\bm{z}_t^s = \mathcal{F}_{\text{comp}}^s( \bm{y}_t^s, R_t^s)$,
where $\mathcal{F}_{\text{comp}}^s(\cdot)$ denotes the compression encoder for a given number of compression bits $R_t^s$. It can be implemented using classical signal processing methods (e.g., feature extraction and quantization) or learning-based compression methods~\cite{berahmand2024autoencoders}. The fusion center receives $\bm{z}_t^s$ and reconstructs an estimate of the original observation: $\hat{\bm{y}}_t^s = \mathcal{F}_{\text{rec}}^s( \bm{z}_t^s)$,
where $\mathcal{F}_{\text{rec}}^s(\cdot)$ is the reconstruction model. The reconstructed observation $\hat{\bm{y}}_t^s$ approximates $\bm{y}_t^s$, but may include distortion due to compression. This distortion directly impacts both the accuracy of state estimation and the quality of control tasks. Therefore, the design of compression algorithms tailored to different objectives in closed-loop \ac{scc} systems can lead to different impacts on overall system performance. This highlights the importance of \emph{goal-oriented semantic communication}, aiming to maintain robust overall system performance while reducing communication overhead through the selective acquisition of task-relevant information.

\subsection{Digital Twin and State Tracker}

The fusion center maintains a real-time model of the tracked objects in the form of a DT, which outputs an estimated state $\hat{\bm{x}}_t$. Due to the bandwidth limitations of the communication links, the \ac{dt} operates on reconstructed sensor observations $\hat{\bm{y}}_{t}^{1:S}$ which are lossy approximations of the original raw data $\bm{y}_{t}^{1:S}$. To support state estimation, the \ac{dt} maintains the following two probabilistic models: 
\begin{itemize}
    \item An observation model $p_{\text{DT}}(\hat{\bm{y}}_{t}^{1:S}|\bm{x}_{t})$, capturing the statistical relationship between the reconstructed observations and the estimated state;

    \item A kinematic prediction model $p_{\text{DT}}(\bm{x}_{t+k} \mid \bm{x}_{t}, \bm{u}_{t})$, which enables multi-step state forecasting for any $k\ge 1$.\footnote{We assume that the \ac{dt} has access to the control inputs $\bm{u}_t^{\text{unctl}}$ of the uncontrollable objects (e.g., through known behavioral models, environmental priors, or direct uploads from the objects themselves). When such information is unavailable, the \ac{dt} can estimate $\bm{u}_t^{\text{unctl}}$ based on historical motion patterns or learned dynamics models, while the remaining uncertainty in these estimates is captured by the process noise in the corresponding state dimensions.} Note that this model may differ from the true dynamics $p_{\text{obj}}(\bm{x}_{t+k} \mid \bm{x}_{t}, \bm{u}_{t})$, depending on the fidelity of the DT. 

\end{itemize}

Let $q_{\text{st}}(\bm{x}_t | \hat{\bm{y}}_{1:t}^{1:S}, \bm{u}_{1:t-1})$ denote the belief over the state at time $t$. Upon receiving reconstructed observations $\hat{\bm{y}}_{t+1}^{1:S}$ and the applied control $\bm{u}_t$, the belief is updated as:
\begin{align}
\label{eq:state_tracker_hat_en}
q_{\text{st}}({\bm{x}}_{t+1} | \hat{\bm{y}}_{1:t+1}^{1:S} , \bm{u}_{1:t})
& \! \propto \!  p_{\text{DT}}(\hat{\bm{y}}_{t+1}^{1:S} | {\bm{x}}_{t+1}) \\
&\! \! \! \! \! \! \! \! \! \! \! \! \times \!\! \int \!\! q_{\text{st}}({\bm{x}}_t | \hat{\bm{y}}_{1:t}^{1:S}, \bm{u}_{1:t-1}) p_{\text{DT}}({\bm{x}}_{t+1} | {\bm{x}}_t, \bm{u}_t) \mathrm{d} {\bm{x}}_t.\nonumber
\end{align}
This belief represents the fusion center’s probabilistic inference of the object state based on compressed and reconstructed observations. If the \ac{dt} models accurately reflect the real system, the belief approximates the true posterior distribution of $\bm{x}_{t+1}$. However, if there are mismatches in the model (e.g., due to compression or incorrect DT), the belief may not accurately capture the posterior. 
The \ac{dt} could be designed to mitigate such errors by aggregating multi-sensor information, modeling reconstruction uncertainty, and exploiting temporal dynamics~\cite{Chen_DigitalTwin_2024, Khan_DigitalTwin_2022}.

\subsection{Controller}
In the controller, the control input $\bm{u}_t^{\text{ctl}}$ is generated according to a policy that optimizes the overall objective of the closed-loop system. This objective typically depends on the system state, control input, and possibly the observations. A general form of the average cumulative cost over a finite horizon $T$ is given by
\begin{align}
\label{eq:cost_function}
J = \frac{1}{T}\sum_{t=1}^{T} J_t(\bm{x}_t, \bm{u}_{t}^{\text{ctl}}, \bm{y}_t^{1:S}),
\end{align}
where $J_t(\bm{x}_t, \bm{u}_{t}^{\text{ctl}}, \bm{y}_t^{1:S})$ denotes the cost for time $t$. To minimize the cost, the controller leverages the belief distribution produced by the \ac{dt} and selects the control action according to
$\bm{u}_{t}^{\text{ctl}} \sim \pi\big(q_{\text{st}}({\bm{x}}_t \mid \hat{\bm{y}}_{1:t}^{1:S}, \bm{u}_{1:t-1})\big) $,
where $\pi(\cdot)$ denotes the control policy. 
This policy can be pre-defined or learned, depending on the system structure and optimization goals. Representative methods include model predictive control (MPC) and RL~\cite{RLMPC}. The generated control actions aim to achieve specific operational objectives, such as guiding vehicles to target positions or stabilizing unmanned aerial vehicle (UAV) trajectories. Finally, these actions are transmitted to the controlled objects, inducing state transitions governed by the dynamics:
$\bm{x}_{t+1} \sim p_{\text{obj}}(\bm{x}_{t+1} \mid \bm{x}_{t}, \bm{u}_{t})$,
thereby closing the feedback loop.

\section{Semantic-Aware Framework for Closed-Loop Distributed \ac{scc} Systems} \label{sec:error-levels}

Semantic communication is an emerging paradigm that emphasizes the meaning and impact of transmitted information, beyond conventional symbol-level accuracy. Nearly 70 years ago, Weaver, in his commentary on Shannon's seminal work, introduced a three-level framework for communication~\cite{shannon1998mathematical}:
\begin{itemize}
    \item \textbf{L1}: How accurately can the symbols of communication be transmitted?
    \item \textbf{L2}: How precisely do the transmitted symbols convey the desired meaning?
    \item \textbf{L3}: How effectively does the received meaning affect conduct in the desired way?
\end{itemize}

This perspective naturally corresponds to the three core components of closed-loop \ac{scc} systems: communication, sensing, and control. It enables a principled, task-oriented design framework for semantic-aware \ac{scc} systems. In the following, we introduce a unified problem formulation for goal-oriented semantic communication in closed-loop distributed \ac{scc} systems by examining these three levels through the lenses of observation reconstruction, state estimation, and control. 
Subsequently, we propose a unified goal-oriented semantic compression and rate adaptation framework applicable to different semantic error levels\footnote{Strictly speaking, the L1 is reconstruction-oriented rather than semantic-based. However, for simplicity, we collectively refer to all three levels as semantic-based in the following.} and their corresponding objectives across the \ac{scc} loop.

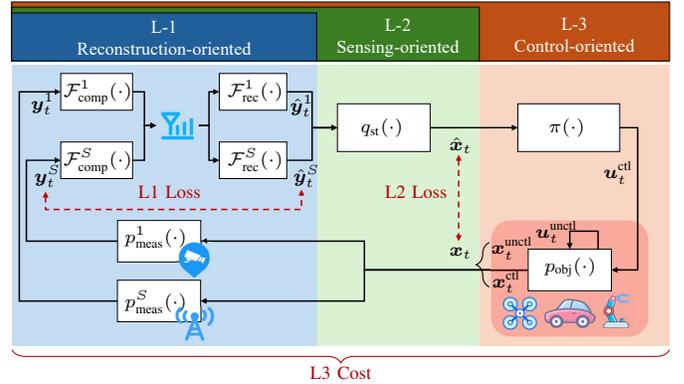
\begin{figure}[t]
    \centering
    \begin{tikzpicture}
    \node (image) [anchor=south west]{\includegraphics[width=1.0\linewidth]{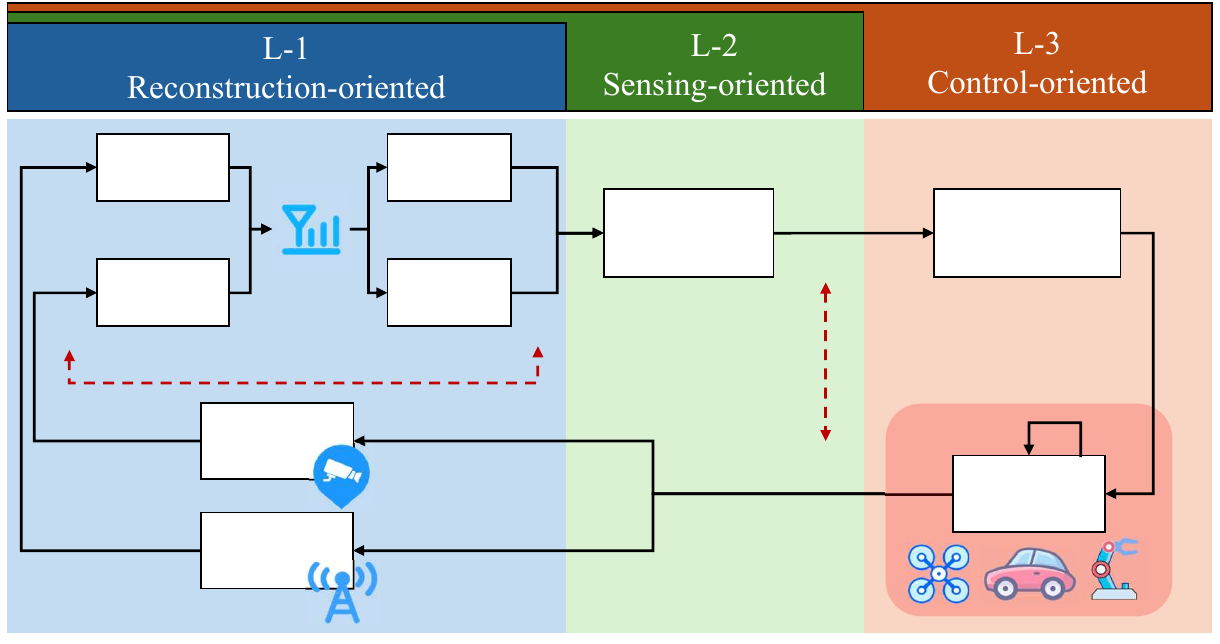}};
    \gettikzxy{(image.north east)}{\ix}{\iy};

     \node at (0.145*\ix,0.785*\iy)[rotate=0,anchor=north]{{\scriptsize $\mathcal{F}_{\text{comp}}^1(\cdot)$}};
     \node at (0.145*\ix,0.6*\iy)[rotate=0,anchor=north]{{\scriptsize $\mathcal{F}_{\text{comp}}^S(\cdot)$}};

    \node at (0.375*\ix,0.785*\iy)[rotate=0,anchor=north]{{\scriptsize $\mathcal{F}_{\text{rec}}^1(\cdot)$}};
     \node at (0.375*\ix,0.6*\iy)[rotate=0,anchor=north]{{\scriptsize $\mathcal{F}_{\text{rec}}^S(\cdot)$}};

     \node at (0.56*\ix,0.68*\iy)[rotate=0,anchor=north]{{\scriptsize $q_{\text{st}}(\cdot)$}};

     \node at (0.83*\ix,0.68*\iy)[rotate=0,anchor=north]{{\scriptsize $\pi(\cdot)$}};

     \node at (0.835*\ix,0.305*\iy)[rotate=0,anchor=north]{{\scriptsize $p_{\text{obj}}(\cdot)$}};

     \node at (0.23*\ix,0.39*\iy)[rotate=0,anchor=north]{{\scriptsize $p_{\text{meas}}^1(\cdot)$}};
     \node at (0.23*\ix,0.215*\iy)[rotate=0,anchor=north]{{\scriptsize $p_{\text{meas}}^S(\cdot)$}};
    
    \node at (0.065*\ix,0.75*\iy)[rotate=0,anchor=north]{{\scriptsize $\bm{y}_t^1$}};
    \node at (0.072*\ix,0.555*\iy)[rotate=0,anchor=north]{{\scriptsize $\bm{y}_t^S$}};

    \node at (0.443*\ix,0.744*\iy)[rotate=0,anchor=north]{{\scriptsize $\hat{\bm{y}}_t^1$}};
    \node at (0.45*\ix,0.555*\iy)[rotate=0,anchor=north]{{\scriptsize $\hat{\bm{y}}_t^S$}};

    \node at (0.675*\ix,0.63*\iy)[rotate=0,anchor=north]{{\scriptsize $\hat{\bm{x}}_t$}};

    \node at (0.905*\ix,0.565*\iy)[rotate=0,anchor=north]{{\scriptsize $\bm{u}_t^{\text{ctl}}$}};
    \node at (0.815*\ix,0.405*\iy)[rotate=0,anchor=north]{{\scriptsize $\bm{u}_t^{\text{unctl}}$}};

    \node at (0.675*\ix,0.331*\iy)[rotate=0,anchor=north]{{\scriptsize ${\bm{x}}_t$}};
    
    \node at (0.742*\ix,0.27*\iy)[rotate=0,anchor=north]{{\scriptsize $\bm{x}_t^{\text{ctl}}$}};
    \node at (0.75*\ix,0.36*\iy)[rotate=0,anchor=north]{{\scriptsize $\bm{x}_t^{\text{unctl}}$}};

    \node at (0.25*\ix,0.5*\iy)[rotate=0,anchor=north] {\color{darkred} {\scriptsize L1 Loss}};
    \node at (0.61*\ix,0.5*\iy)[rotate=0,anchor=north] {\color{darkred} {\scriptsize L2 Loss}};
    \node at (0.5*\ix,0.01*\iy)[rotate=0,anchor=north] {\color{darkred} {\scriptsize L3 Cost}};

    \draw[decorate, decoration={brace, amplitude=5pt}] 
    (0.717*\ix,0.20*\iy) -- (0.717*\ix,0.32*\iy) %
    node[midway, right=3pt, yshift=0.5mm]{};

    \draw[darkred, decorate, decoration={brace, amplitude=5pt, mirror}]
        (0.02*\ix,0.03*\iy) -- (0.98*\ix,0.03*\iy)
        node[midway, above=3pt]{};
    
    \end{tikzpicture}
    \vspace{-4mm}
    \caption{Schematic diagram of three semantic levels for distributed closed-loop \ac{scc} systems.}

    \label{fig:errorlevels}
    \vspace{-1mm}
\end{figure}

\subsection{Semantic-Aware Problem Formulation}
\label{sec:problem}

To evaluate the impact of goal-oriented semantic communication in \ac{scc} systems, we formulate optimization problems corresponding to the three levels of semantic error. As shown in Fig. \ref{fig:errorlevels}, each level reflects a distinct stage in the closed-loop pipeline (i.e., reconstruction, estimation, and control cost), and jointly considers semantic compression and rate adaptation.

\subsubsection{Reconstruction-Level Error (L1)} At L1, the objective is to minimize the distortion between the original and reconstructed sensor observations. The encoder, decoder, and transmission rate for each sensor are jointly optimized to ensure accurate data reconstruction, without considering downstream tasks such as estimation or control. Let $\bm{\theta}_{\text{comp}}^s$ and $\bm{\theta}_{\text{rec}}^s$ denote the parameters of the encoder and decoder for sensor~$s$, respectively, and let 
$\mathcal{F}_{\text{rate}}(\cdot;\bm{\theta}_{\text{rate}})$ be the rate selection policy with parameter~$\bm{\theta}_{\text{rate}}$. 
The function $\ell(\bm{y}_t^s,\hat{\bm{y}}_t^s)$ represents a distortion measure (e.g., mean squared error) 
between the original observation $\bm{y}_t^s$ and its reconstruction $\hat{\bm{y}}_t^s$. 
The optimization problem is formulated as follows:
\begin{Prob}[Reconstruction-Oriented Communication Problem]
\label{Prob:L1} At L1, the encoders, decoders, and rate selection are determined to solve the following problem:
\begin{align}
\min_{\substack{
   \{\bm{\theta}_{\text{comp}}^s , \bm{\theta}_{\text{rec}}^s\}_{s=1}^S, \\
   \bm{\theta}_{\text{rate}}
}}
&\quad\frac{1}{T}\sum_{t=1}^{T}\!\sum_{s=1}^S \ell\!\left( \bm{y}_t^s, \hat{\bm{y}}_t^s \right)  \label{eq:P1} \\
\text{\rm s.t.}\quad 
& \ \{R_t^s\}_{s=1}^S = \mathcal{F}_{\text{rate}}\!\big(\bm{U}_{1,t}; \bm{\theta}_{\text{rate}}), \tag{\ref{eq:P1}a} \label{eq:P1-a}\\
& \ \bm{z}_t^s = \mathcal{F}_{\text{comp}}^s\!\left( \bm{y}_t^s, R_t^s;\, \bm{\theta}_{\text{comp}}^s\right), \ \ \forall s, \tag{\ref{eq:P1}b} \label{eq:P1-b}  \\
& \ \hat{\bm{y}}_t^s = \mathcal{F}_{\text{rec}}^s\!\left( \bm{z}_t^s;\, \bm{\theta}_{\text{rec}}^s\right), \ \ \forall s, \tag{\ref{eq:P1}c}  \label{eq:P1-c}\\
& \frac{1}{T}\sum_{t=1}^T \sum_{s\in \mathcal{S}} R_t^s \le R  \tag{\ref{eq:P1}d} \label{eq:P1-d}.
\end{align}
\end{Prob} 
\noindent Here $\bm{U}_{1,t}$ consists of communication-related information. 
For example, for wireless sensing BSs, $\bm{U}_{1,t}$ includes the current channel conditions and noise power, which collectively reflect the quality of observation reconstruction.

\subsubsection{Sensing-Level Error (L2)} Building upon L1, the goal at L2 is to minimize the state estimation error at the fusion center. At L2, the encoder and decoder are thus optimized to retain information most relevant to accurate state estimation under rate constraints. Let $\ell(\bm{x}_t,\hat{\bm{x}}_t)$ denote a state-level distortion measure between the true state $\bm{x}_t$ and its estimation $\hat{\bm{x}}_t$. The optimization problem is formulated as follows:
\begin{Prob}[Sensing-Oriented Communication Problem]
\label{Prob:L2}At L2, the encoders, decoders, and rate selection are determined to solve the following problem:
\begin{align}
\min_{\substack{
   \{\bm{\theta}_{\text{comp}}^s , \bm{\theta}_{\text{rec}}^s\}_{s=1}^S, \\
   \bm{\theta}_{\text{rate}}
}} & \quad\frac{1}{T}\sum_{t=1}^{T}  \ell\!\left(\bm{x}_t, \hat{\bm{x}}_t \right)  \label{eq:P2} \\
\text{s.t.}\quad 
& \ \{R_t^s\}_{s=1}^S = \mathcal{F}_{\text{rate}}\!\big(\bm{U}_{2,t}; \bm{\theta}_{\text{rate}}),  \tag{\ref{eq:P2}a} \label{eq:P2-a}\\
& \ \hat{\bm{x}}_t \sim q_{\text{st}}({\bm{x}}_t \mid \hat{\bm{y}}_{1:t}^{1:S}, \bm{u}_{1:t-1}), \tag{\ref{eq:P2}b} \label{eq:P2-b}\\
& \eqref{eq:P1-b}, \eqref{eq:P1-c}, \eqref{eq:P1-d}. \nonumber
\end{align}
\end{Prob}
\noindent Building upon L1, $\bm{U}_{2,t}$ consists of sensing- and estimation-related indicators. 
In addition to the information contained in $\bm{U}_{1,t}$, it also includes measures of the confidence in the system state, such as the state tracker's belief distribution $q_{\text{st}}(\bm{x}_t)$ and historical estimated states.

\subsubsection{Control-Level Error (L3)} At L3, the objective is to minimize the ultimate control cost, capturing how semantic compression influences the overall performance of the closed-loop system. This formulation closes the semantic loop by directly linking data transmission to task-specific goals such as trajectory tracking accuracy, control effort, or system stability. The L3 optimization problem can be formulated as follows:
\begin{Prob}[Control-Oriented Communication Problem]
\label{Prob:L3}At L3, the encoders, decoders, and rate selection are determined to solve the following problem:
\begin{align}
\min_{\substack{
   \{\bm{\theta}_{\text{comp}}^s , \bm{\theta}_{\text{rec}}^s\}_{s=1}^S, \\
   \bm{\theta}_{\text{rate}}
   }}
& \quad\frac{1}{T}\sum_{t=1}^{T} J_t(\bm{x}_t, \bm{u}_t^{\text{ctl}}, \bm{y}_t^{1:S}) \label{eq:P3} \\
\text{s.t.}\quad 
& \ \{R_t^s\}_{s=1}^S = \mathcal{F}_{\text{rate}}\!\big(\bm{U}_{3,t}; \bm{\theta}_{\text{rate}}), \tag{\ref{eq:P3}a}\\
& \ \bm{u}_t \sim \pi\!\left( q_{\text{st}}({\bm{x}}_t \mid \hat{\bm{y}}_{1:t}^{1:S}, \bm{u}_{1:t-1}) \right), \tag{\ref{eq:P3}b}\\
& \eqref{eq:P1-b}, \eqref{eq:P1-c}, \eqref{eq:P1-d} , \eqref{eq:P2-b}. \nonumber
\end{align}
\end{Prob}
In addition to the information included in L1 and L2, $\bm{U}_{3,t}$ further incorporates control-related features, such as the target state and historical control commands.

\begin{figure}[t]
    \centering
    \begin{tikzpicture}
    \node (image) [anchor=south west]{\includegraphics[width=1.0\linewidth]{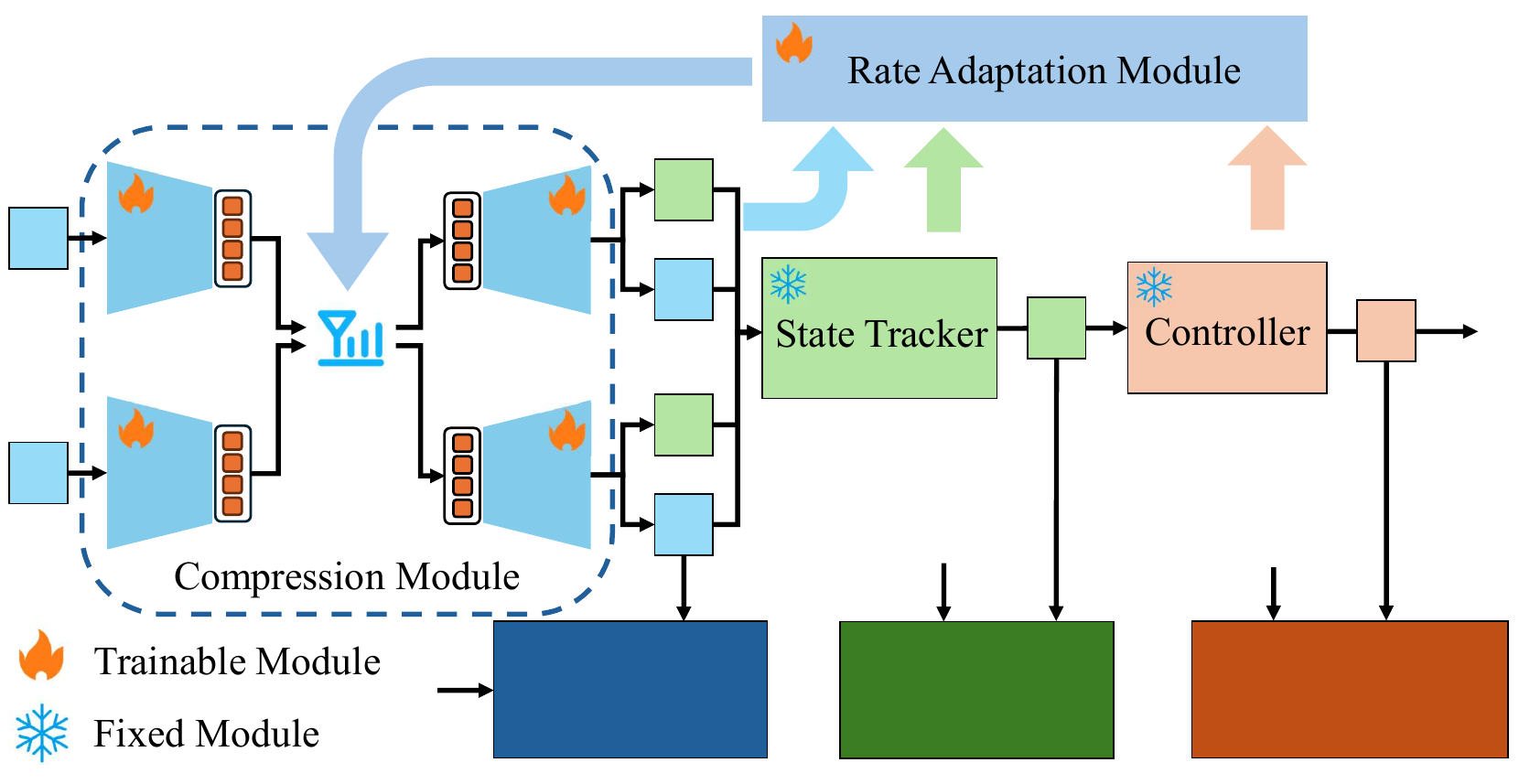}};
    \gettikzxy{(image.north east)}{\ix}{\iy};
    
    \node at (0.04*\ix,0.742*\iy)[rotate=0,anchor=north]{{\scriptsize $\bm{y}_t^S$}};
    \node at (0.04*\ix,0.454*\iy)[rotate=0,anchor=north]{{\scriptsize $\bm{y}_t^1$}};    

     \node at (0.115*\ix,0.73*\iy)[rotate=0,anchor=north]{{\scriptsize $\bm{\theta}_{\text{comp}}^S$}};
     \node at (0.115*\ix,0.44*\iy)[rotate=0,anchor=north]{{\scriptsize $\bm{\theta}_{\text{comp}}^1$}};
    \node at (0.35*\ix,0.44*\iy)[rotate=0,anchor=north]{{\scriptsize $\bm{\theta}_{\text{rec}}^1$}};
    \node at (0.35*\ix,0.73*\iy)[rotate=0,anchor=north]{{\scriptsize $\bm{\theta}_{\text{rec}}^S$}};

    \node at (0.55*\ix,0.795*\iy)[rotate=0,anchor=north]{{\scriptsize $\bm{U}_{1,t}^s$}};
    \node at (0.625*\ix,0.79*\iy)[rotate=0,anchor=north]{{\scriptsize $\bm{U}_{2,t}$}};
    \node at (0.835*\ix,0.79*\iy)[rotate=0,anchor=north]{{\scriptsize $\bm{U}_{3,t}$}};

    \node at (0.47*\ix,0.993*\iy)[rotate=0,anchor=north]{{\scriptsize ${R}_{t}^s$}};

    \node at (0.454*\ix,0.39*\iy)[rotate=0,anchor=north]{{\scriptsize $\hat{\bm{y}}_t^1$}};
    \node at (0.454*\ix,0.68*\iy)[rotate=0,anchor=north]{{\scriptsize $\hat{\bm{y}}_t^S$}};
    \node at (0.455*\ix,0.514*\iy)[rotate=0,anchor=north]{{\scriptsize ${\bm{\zeta}}_t^1$}};
    \node at (0.454*\ix,0.802*\iy)[rotate=0,anchor=north]{{\scriptsize ${\bm{\zeta}}_t^S$}};    

    \node at (0.692*\ix,0.63*\iy)[rotate=0,anchor=north]{{\scriptsize $\hat{\bm{x}}_t$}};

    \node at (0.902*\ix,0.612*\iy)[rotate=0,anchor=north]{{\scriptsize $\bm{u}_t$}};

    \node at (0.305*\ix,0.222*\iy)[rotate=0,anchor=north]{{\scriptsize \scriptsize ${\bm{y}}_t^s$}};

    \node at (0.60*\ix,0.30*\iy)[rotate=0,anchor=north]{{\scriptsize \scriptsize ${\bm{x}}_t$}};

    \node at (0.42*\ix,0.22*\iy)[rotate=0,anchor=north]{{\scriptsize \textcolor{white}{L1 Loss}}}; 
    \node at (0.42*\ix,0.15*\iy)[rotate=0,anchor=north]{{\scriptsize \textcolor{white}{$\ell\left( \bm{y}_t^s, \hat{\bm{y}}_t^s \right)$}}};

    \node at (0.644*\ix,0.22*\iy)[rotate=0,anchor=north]{{\scriptsize \textcolor{white}{L2 Loss}}};    
    \node at (0.644*\ix,0.15*\iy)[rotate=0,anchor=north]{{\scriptsize \textcolor{white}{$\ell\left( \bm{x}_t, \hat{\bm{x}}_t \right)$}}};
    
    \node at (0.88*\ix,0.22*\iy)[rotate=0,anchor=north]{{\scriptsize \textcolor{white}{L3 Loss}}};
    \node at (0.88*\ix,0.15*\iy)[rotate=0,anchor=north]{{\scriptsize \textcolor{white}{$J_t(\bm{x}_t,\bm{u}_t,\bm{y}_t^s)$}}};
    \node at (0.81*\ix,0.30*\iy)[rotate=0,anchor=north]{{\scriptsize \scriptsize $\bm{x}_t$}};
    \node at (0.855*\ix,0.316*\iy)[rotate=0,anchor=north]{{\scriptsize \scriptsize $\bm{y}_t^s$}};

    \end{tikzpicture}
    \caption{Schematic diagram of the general framework for semantic-aware communication and rate adaptation in distributed closed-loop \ac{scc} systems. It illustrates a unified framework encompassing three semantic levels. For different semantic levels, we could activate different loss functions.}
    
    \label{fig:Unified-Framework}
\end{figure}
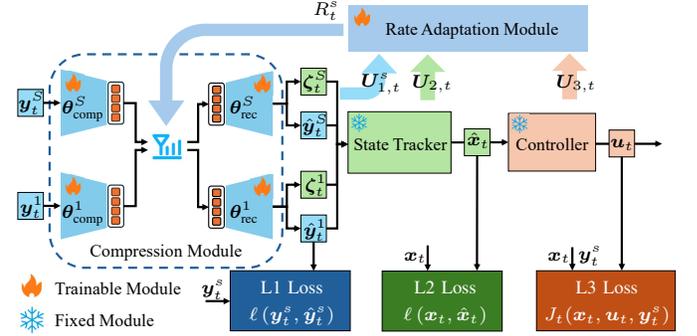

\subsection{Proposed Framework for Three-Level Problems}
\label{sec:Unified-Framework}

Based on the above problem formulations, we introduce a unified framework for semantic-aware communication and adaptation in distributed closed-loop \ac{scc} systems. As shown in Fig.~\ref{fig:Unified-Framework},
the framework integrates \textit{semantic compression module} and \textit{rate adaptation module}, aiming to jointly optimize sensing, communication, and control performance. In the following, we introduce the framework in detail.

\subsubsection{Semantic Compression Module}
Semantic compression module learns encoder/decoder parameters $\{\bm{\theta}_{\text{comp}}^s,\bm{\theta}_{\text{rec}}^s\}_{s=1}^S$ under different rate allocation policies to optimize the level-specific objective i.e.,  observation reconstruction (L1), state estimation (L2), or control cost (L3).

\textbf{L1 Semantic Compression:}
For L1, the objective is to reconstruct the original observation as faithfully as possible. Given a fixed compression rate $R_t^s$, we can temporarily ignore constraints~\eqref{eq:P1-a} and ~\eqref{eq:P1-d}. The encoder and decoder parameters are optimized to minimize the reconstruction distortion:
\begin{align}
\min_{ \bm{\theta}_{\text{comp}}^s , \bm{\theta}_{\text{rec}}^s } 
\quad \mathcal{L}_1 =\frac{1}{T}\sum_{t=1}^{T} \ell\!\left( \bm{y}_t^s, \hat{\bm{y}}_t^s \right).
\end{align}
In this case, training is straightforward: the encoder input $\bm{y}_t^s$ serves as the ground truth for the decoder’s output $\hat{\bm{y}}_t^s$.

\textbf{L2 Semantic Compression:}
When considering L2, the goal shifts from observation reconstruction to accurate state estimation at the fusion center:
\begin{align}
\min_{ \bm{\theta}_{\text{comp}}^s , \bm{\theta}_{\text{rec}}^s } 
\quad \mathcal{L}_2 =\frac{1}{T}\sum_{t=1}^{T} \ell\!\left( \bm{x}_t, \hat{\bm{x}}_t \right) .
\end{align}
Compared with L1, two additional aspects arise. 
 Firstly, the DT-based state tracker 
$q_{\text{st}}(\cdot)$ must be taken into account.  
    For a differentiable tracker, it can be directly embedded into the training loop so that gradients flow back to the encoder and decoder. For a non-differentiable tracker, an imitator model~\cite{zare2024survey} can be trained to approximate it, enabling gradient-based optimization while using the original tracker at inference. Secondly, 
    due to compression-induced distortion, the original observation model $p_{\text{DT}}(\bm{y}_t|\bm{x}_t)$ may become mismatched. To compensate, the semantic decoder outputs both the reconstructed observation $\hat{\bm{y}}_t$ and an uncertainty estimate $\bm{\zeta}_t$, enabling the state tracker to adaptively weigh uncertain measurements and enhance estimation robustness. 

\textbf{L3 Semantic Compression:}
At L3, the optimization directly targets the closed-loop control objective. 
Instead of reconstruction or estimation error, the training loss is the task-specific control cost:
\vspace{-1mm}
\begin{align}
\min_{ \bm{\theta}_{\mathrm{comp}}^s , \bm{\theta}_{\mathrm{rec}}^s }
\quad \mathcal{L}_3 =\frac{1}{T}\sum_{t=1}^{T} J_t\!\left( {\bm{x}}_t, \bm{u}_{t}^{\text{ctl}}, {\bm{y}}_t^{1:S} \right),
\end{align}
During encoder–decoder training, the control policy $\pi(\cdot)$ is kept fixed. 
It should also be differentiable so that gradients can propagate through the entire chain of compression, estimation, and control. For non-differentiable controllers, imitation learning~\cite{zare2024survey} can also approximate the control policy to enable gradient-based training.
This closes the semantic loop by aligning communication strategies with the final operational goals of the closed-loop \ac{scc} system.

\subsubsection{Goal-Oriented Rate Adaptation Module}
Rate adaptation module adaptively select per-sensor rates $\{R_t^s\}_{s=1}^S$ over time to minimize the chosen loss/cost while satisfying the rate constraint $\sum_{s=1}^S R_t^s \le R_t$. The optimization problem is formulated as
\vspace{-1mm}
\begin{align}
\label{Prob:RateAdapt}
\min_{\bm{\theta}_{\text{rate}}} 
& \quad \mathcal{L}_l \\
\text{s.t.}\quad 
& \ \{R_t^s\}_{s=1}^S  
    = \mathcal{F}_{\text{rate}}\!\big(\bm{U}_{l,t};\,\bm{\theta}_{\text{rate}}\big), \label{eq:rate_mapping_constraint}\\
& \ \frac{1}{T}\sum_{t=1}^T \sum_{s\in \mathcal{S}} R_t^s \le R , \label{eq:rate_budget_constraint}
\end{align}
where $\mathcal{L}_l$ $(l=1,2,3)$ denotes the semantic-level objective introduced in the semantic compression module. 
The function $\mathcal{F}_{\text{rate}}(\cdot)$ is a parameterized mapping with learnable parameters $\bm{\theta}_{\text{rate}}$ shared among all sensors. In \eqref{eq:rate_mapping_constraint}, 
$\bm{U}_{l,t}$ represents the system information under level~$l$, which may correspond to either an abstract latent representation or a physically interpretable variable, serving as the informational basis for adaptive rate allocation.
This formulation enables the system to coordinate compression rates across sensors for different semantic error consideration, while satisfying the rate constraint~\eqref{eq:rate_budget_constraint}. The design of the input feature $\bm{U}_{l,t}$ depends on the level under consideration, as introduced in Sec. \ref{sec:problem}.
These level-dependent context features serve as the informational basis for adaptive rate allocation in the proposed framework. In practice, $\mathcal{F}_{\text{rate}}(\cdot)$ can be implemented as a neural network or a lightweight decision rule, trained jointly with the encoder–decoder (for L1/L2) or together with the control policy (for L3).

\section{Case Study: Rate-limited \ac{lqr} System}
To demonstrate the applicability of the proposed framework, we consider a rate-limited distributed \ac{lqr} system as a representative example. 
This case study illustrates how the general distributed \ac{scc} framework can be instantiated and analyzed under communication constraints, demonstrating the impact of semantic compression and dynamic rate adaptation on closed-loop control performance.

\subsection{System Model}
In the \ac{lqr} case study, the environment consists of controllable objects and uncontrollable objects. 
Distributed sensors observe these objects and transmit compressed measurements over rate-limited communication links to a fusion center. 
At the fusion center, a state tracker, implemented via a Kalman filter, produces an estimate of the overall system state. 
Based on this estimate, the controller generates control actions for the controllable object and sends them back to the plant, thereby forming a closed-loop \ac{scc} system.

\subsubsection{State Model}

We assume that the controllable and uncontrollable objects evolve independently, 
such that their state transitions and control inputs do not mutually affect each other. 
Accordingly, the overall system dynamics can be written as
\begin{align} \label{eq:dynamics}
    \!\! \! \bm{x}_{t+1} \! = \!
    \underbrace{
    \begin{bmatrix}
        \bm{A}^{\text{ctl}} & \bm{0} \\
        \bm{0} & \bm{A}^{\text{unctl}}
    \end{bmatrix}}_{\bm{A}}
    \underbrace{
    \begin{bmatrix}
        \bm{x}_t^{\text{ctl}} \\
        \bm{x}_t^{\text{unctl}}
    \end{bmatrix}}_{\bm{x}_t}
    +
    \underbrace{
    \begin{bmatrix}
        \bm{B}^{\text{ctl}} & \bm{0} \\
        \bm{0} &\bm{B}^{\text{unctl}}
    \end{bmatrix}}_{\bm{B}}
    \underbrace{
    \begin{bmatrix}
        \bm{u}_t^{\text{ctl}} \\
        \bm{u}_t^{\text{unctl}}
    \end{bmatrix}}_{\bm{u}_t}
    + \bm{v}_{t}, \nonumber
\end{align}
where $\bm{x}_t^{\text{ctl}} \in \mathbb{R}^{N_x^{\text{ctl}}}$ and $\bm{x}_t^{\text{unctl}}\in \mathbb{R}^{N_x^{\text{unctl}}}$ denote the states of the controllable and uncontrollable objects, respectively, 
and $\bm{u}_t^{\text{ctl}}\in \mathbb{R}^{N_u^{\text{ctl}}}$ and $\bm{u}_t^{\text{unctl}}\in \mathbb{R}^{N_u^{\text{unctl}}}$ are their corresponding control inputs. 
The matrices $\bm{A}^{\text{ctl}}$ and $\bm{A}^{\text{unctl}}$ describe the internal dynamics of the controllable and uncontrollable objects, respectively, 
while $\bm{B}^{\text{ctl}}$ and $\bm{B}^{\text{unctl}}$ represent their respective control gain matrices. The process noise $\bm{v}_t$ is assumed to be Gaussian, i.e., $\bm{v}_t \sim \mathcal{N}(\bm{0}, \bm{Q})$, with covariance $\bm{Q} \in \mathbb{R}^{(N_x^{\text{ctl}}+N_x^{\text{unctl}}) \times (N_x^{\text{ctl}}+N_x^{\text{unctl}})}$.

\subsubsection{Sensor Model}

At each time step, sensor $s \in \{1, \ldots, S\}$ observes both the controllable and uncontrollable objects through a linear measurement model:
\begin{align}
\bm{y}_{t}^s 
    = 
    \bm{C}^s
    [
(\bm{x}^{\text{ctl}}_t)^\top, 
(\bm{x}^{\text{unctl}}_t)^\top
]^\top + \bm{w}_{t}^s,\end{align}
where $\bm{y}_t^s \in \mathbb{R}^{N_y^s}$ denotes the observation collected by sensor~$s$, 
and $\bm{C}^s \in \mathbb{R}^{N_y^s \times (N_x^{\text{ctl}} + N_x^{\text{unctl}})}$ is the known observation matrix associated with both the controllable and uncontrollable states. 
The noise $\bm{w}_t^s$ is modeled as zero-mean Gaussian, i.e., $\bm{w}_t^s \sim \mathcal{N}(\bm{0}, \bm{W}^s_t)$, 
with covariance $\bm{W}^s_t \in \mathbb{R}^{N_y^s \times N_y^s}$.

\subsubsection{Rate-limited Links}

Due to bandwidth constraints, the raw observations $\bm{y}_t^s$ must be compressed before transmission. 
Each sensor encodes its high-dimensional measurement into a compact representation $\tilde{\bm{z}}_t^s \in \mathbb{R}^{N_z^s}$ using an encoder $\mathcal{F}_{\text{comp}}^s(\bm{y}_t^s, N_z^s; \bm{\theta}_{\text{comp}}^s)$, where $N_z^s$ denotes the compression dimension.
To adapt the communication cost to the sensing quality of each sensor (e.g., its observation noise level $\bm{W}^s_t$), 
the system performs dynamic quantization of the compressed representation. 
Specifically, each element of $\tilde{\bm{z}}_t^s$ is quantized to $r_t^s$ bits, producing the transmitted codeword:
$  \bm{z}_t^s = \mathcal{Q}(\tilde{\bm{z}}_t^s, R_t^s)$,
where $\mathcal{Q}(\cdot, \cdot)$ represents the quantization function and $R_t^s$ is the total number of transmitted bits from sensor~$s$ at time~$t$:
    $R_t^s = N_z^s \, r_t^s$.
We consider a long-term average formulation that reflects the overall communication budget over the entire time horizon~$T$:
$ \frac{1}{T}\sum_{t=1}^T \sum_{s \in \mathcal{S}} R_t^s \le R$,
where $R$ is the long-term communication budget.
After receiving $\bm{z}_t^s$, the fusion center reconstructs the observation as $\hat{\bm{y}}_t^s = \mathcal{F}_{\text{rec}}^s(\bm{z}_t^s; \bm{\theta}_{\text{rec}}^s)$. 
The reconstructed observation $\hat{\bm{y}}_t^s$ is then forwarded to the fusion center for state estimation and control.

\subsubsection{State Tracker}

Given the reconstructed observations, 
the fusion center employs a Kalman filter, denoted by $\mathcal{F}_{\text{KF}}(\cdot)$, 
to estimate the state $\hat{\bm{x}}_{t}$ and its associated error covariance $\bm{\Sigma}_{t|t}$. 
The estimation process consists of two steps~\cite{khodarahmi2023review}:
\begin{itemize}
    \item \textbf{Prediction}:     Based on the system dynamics, the predicted state and its error covariance are given by
    \begin{align}
        \hat{\bm{x}}_{t|t-1} &= \bm{A} \hat{\bm{x}}_{t-1} + \bm{B} \bm{u}_{t-1}, \\
        \bm{\Sigma}_{t|t-1} &= \bm{A} \bm{\Sigma}_{t-1|t-1} \bm{A}^\top + \bm{Q}_t,
    \end{align}
    where $\hat{\bm{x}}_{t|t-1}$ is the predicted state and $\bm{\Sigma}_{t|t-1}$ is the corresponding prediction covariance.

   \item \textbf{Update}: The sensor observations are then used to correct the predicted state estimate:
\begin{align}
    \bm{K}_{t} &= \bm{\Sigma}_{t|t-1} \bm{C}^\top (\bm{C} \bm{\Sigma}_{t|t-1} \bm{C}^\top + \bm{W}_t)^{-1}, \\
    \hat{\bm{x}}_{t} &= \hat{\bm{x}}_{t|t-1} + \bm{K}_{t} (\hat{\bm{y}}_{t} - \bm{C} \hat{\bm{x}}_{t|t-1}), \\
    \bm{\Sigma}_{t|t} &= (\bm{I} - \bm{K}_{t} \bm{C}) \bm{\Sigma}_{t|t-1},
\end{align}
where $\bm{K}_t$ is the Kalman gain, and $\hat{\bm{y}}_t = [\hat{\bm{y}}_t^1, \cdots, \hat{\bm{y}}_t^S]$ is the concatenated vector of reconstructed observations. $\bm{C} = [\bm{C}^1, \ldots, \bm{C}^S]$ is the concatenated observation matrix, and $\bm{W}_t = \mathrm{diag}(\bm{W}^1_t, \ldots, \bm{W}^S_t)$ is the block-diagonal noise covariance matrix.
\end{itemize}
By iteratively applying the prediction and update steps, $\mathcal{F}_{\text{KF}}(\cdot)$ produces an estimate $\hat{\bm{x}}_{t}$ that is subsequently used for control.

\subsubsection{Controller}

The overall control process consists of two parts. The control input for the controllable object, i.e., $\bm{u}_t^{\text{ctl}}$, 
follows an \ac{lqr} framework to minimize a long-term quadratic cost, 
whereas the control input of the uncontrollable subsystem, $\bm{u}_t^{\text{unctl}}$, 
is autonomously generated by the objects themselves. 
In this case study, our objective is to minimize the \ac{lqr} cost associated with the controllable object.
Let $\bm{x}_t^{\text{ctl},\text{desired}} \in \mathbb{R}^{N_x^\text{ctl}}$ denote the desired state of the controllable objects at time~$t$, 
which may correspond to a fixed target or a time-varying reference trajectory. 
The state deviation is defined as
$\tilde{\bm{x}}_t^{\text{ctl}} = \hat{\bm{x}}_{t}^{\text{ctl}} - \bm{x}_t^{\text{ctl},\text{desired}}$,
where $\hat{\bm{x}}_{t}^{\text{ctl}}$ is the estimated state of controllable target obtained from $\hat{\bm{x}}_{t}$. The instantaneous cost at time~$t$ is given by
\begin{align}
    J_t 
    = ({\tilde{\bm{x}}_t^{\text{ctl}}})^\top \bm{Q}_{\mathrm{goal}} \tilde{\bm{x}}_t^{\text{ctl}}
    + (\bm{u}_t^{\text{ctl}})^\top \bm{R}_{\mathrm{goal}} \bm{u}_t^{\text{ctl}},
\end{align}
where $\bm{Q}_{\mathrm{goal}} \succeq 0$ and $\bm{R}_{\mathrm{goal}} \succ 0$ are weighting matrices 
that penalize the state deviation and control effort of the controllable object, respectively. The infinite-horizon objective is to minimize the cumulative cost, i.e., $J = \sum_{t=1}^{\infty} J_t$.
Under assumptions of stabilizability and detectability, 
the optimal control law for the controllable object is expressed as
$\bm{u}_t^{\text{ctl}} = -\bm{K}_{\text{LQR}}\,\hat{\bm{x}}_{t}^{\text{ctl}}$,
where $\bm{K}_{\text{LQR}}$ is the feedback gain matrix obtained by solving the discrete-time algebraic Riccati equation (DARE)~\cite{shaiju2008formulas}.

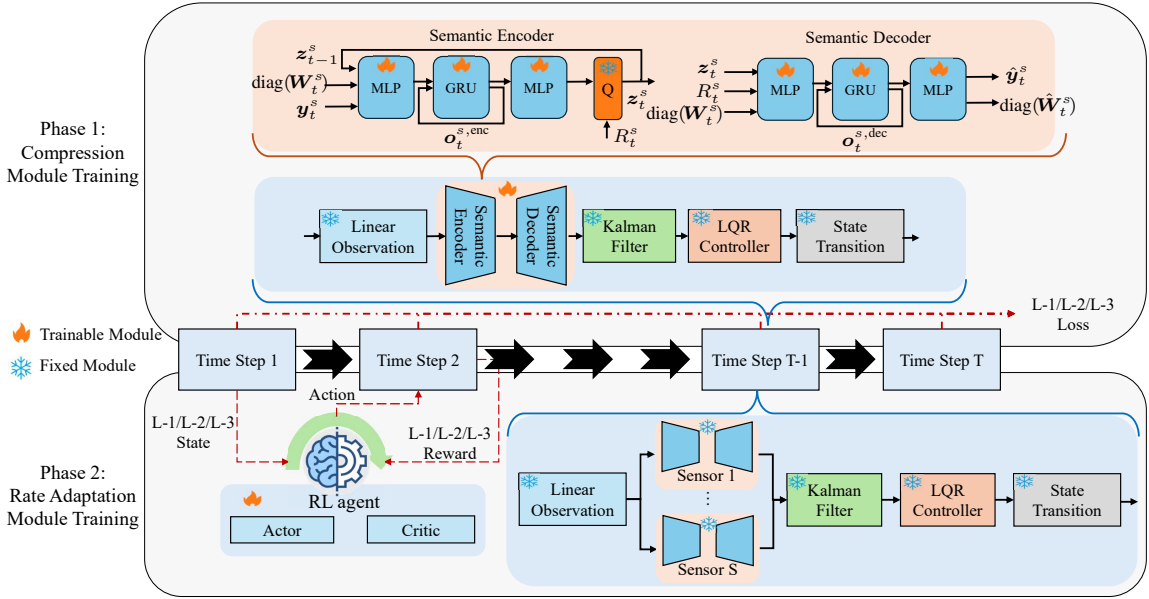
\begin{figure*}[t]
    \centering
    \begin{tikzpicture}
    \node (image) [anchor=south west]{\includegraphics[width=0.85\linewidth]{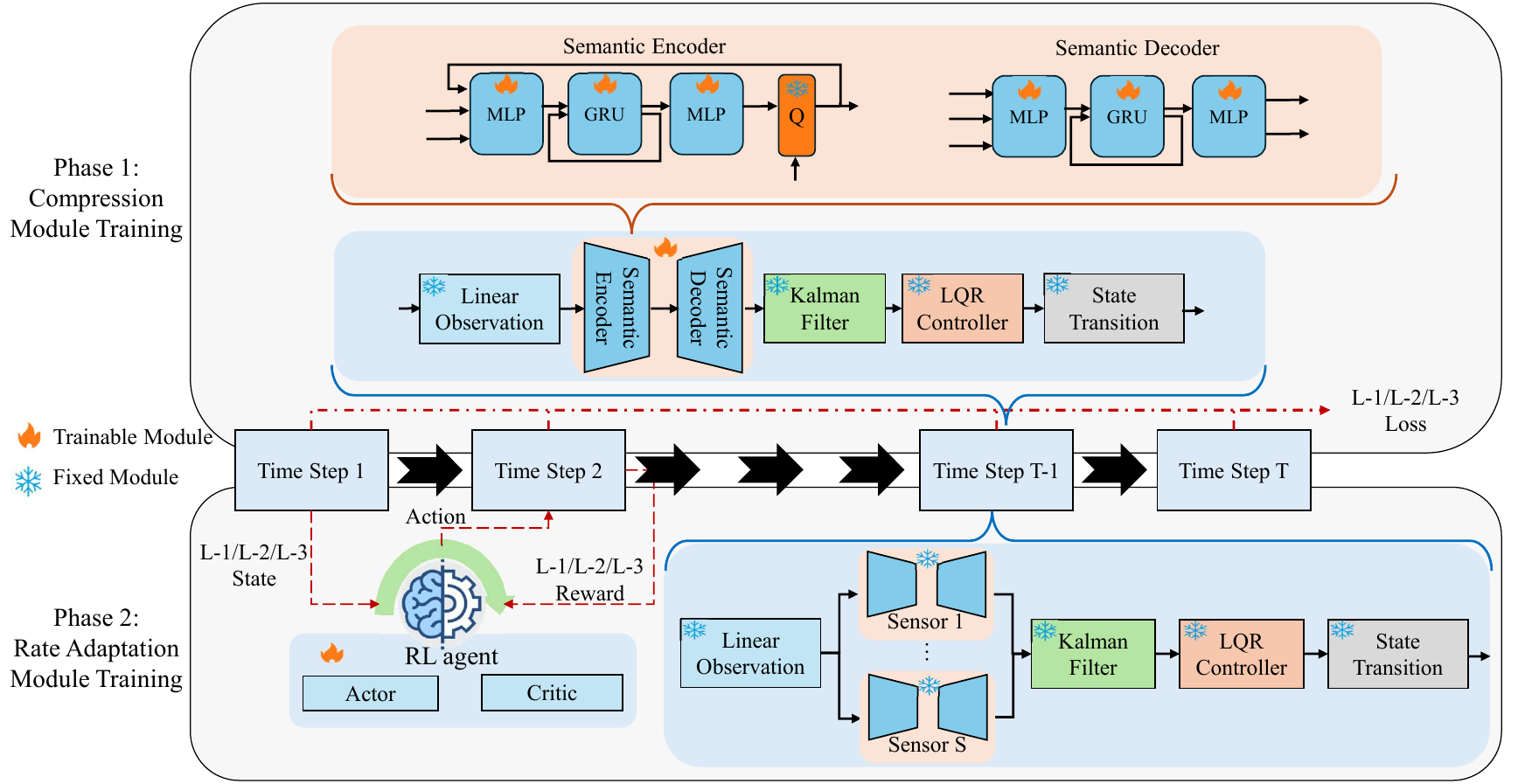}};
    \gettikzxy{(image.north east)}{\ix}{\iy};
    
    \node at (0.27*\ix,0.838*\iy)[rotate=0,anchor=north]{{\scriptsize $\bm{y}_t^s$}};
    \node at (0.255*\ix,0.880*\iy)[rotate=0,anchor=north]{{\scriptsize $\text{diag}(\!\bm{W}^s_t\!)$}};
    \node at (0.278*\ix,0.927*\iy)[rotate=0,anchor=north]{{\scriptsize $\bm{z}_{t-1}^s$}};

    \node at (0.405*\ix,0.793*\iy)[rotate=0,anchor=north]{{\scriptsize $\bm{o}_t^{s,\text{enc}}$}};
    \node at (0.54*\ix,0.793*\iy)[rotate=0,anchor=north]{{\scriptsize $R_{t}^s$}};
    \node at (0.55*\ix,0.855*\iy)[rotate=0,anchor=north]{{\scriptsize $\bm{z}_{t}^s$}};

    \node at (0.61*\ix,0.905*\iy)[rotate=0,anchor=north]{{\scriptsize $\bm{z}_{t}^s$}};
    \node at (0.59*\ix,0.830*\iy)[rotate=0,anchor=north]{{\scriptsize $\text{diag}(\!\bm{W}^s_t\!)$}};  
    \node at (0.61*\ix,0.865*\iy)[rotate=0,anchor=north]{{\scriptsize $R_{t}^s$}};
    \node at (0.743*\ix,0.793*\iy)[rotate=0,anchor=north]{{\scriptsize $\bm{o}_t^{s,\text{dec}}$}};
    \node at (0.872*\ix,0.893*\iy)[rotate=0,anchor=north]{{\scriptsize $\hat{\bm{y}}_t^s$}};
    \node at (0.89*\ix,0.85*\iy)[rotate=0,anchor=north]{{\scriptsize $\text{diag}(\!\hat{\bm{W}}_t^s\!)$}};

    \end{tikzpicture}
    \caption{The proposed semantic communication framework is designed for closed-loop distributed \ac{lqr} systems.  
It consists of a semantic compression module and a rate adaptation module, which are trained sequentially to achieve goal-oriented performance under communication constraints.
}
    
    \label{fig:LQRframework}
    \vspace{-5mm}
\end{figure*}

\subsection{Implementation in the \ac{lqr} Case Study}
\label{subsec:lqrImplementation}

In this subsection, we present the implementation details of the proposed semantic compression and rate adaptation algorithms in the \ac{lqr} case study. 
To ensure a fair comparison with the \ac{lqr} baseline without rate constraints, we retain the Kalman filter for state estimation and the \ac{lqr} controller for control action generation. This setup also accommodates a restricted setting in which the estimation and control components are assumed to be optimally designed, fixed, and known. Under this assumption, improving the communication module alone can still lead to improved overall system performance. 
Semantic compression and rate adaptation are then applied under the three semantic levels described earlier, each corresponding to a different optimization target within the closed-loop system. Specifically, in the \ac{lqr} setting, we consider the following three semantic error levels:
\begin{itemize}
\item \textbf{L1 (Reconstruction Level):} 
The objective is to minimize mean squared error (MSE) of each sensor’s observation. 
The optimization problem is formulated as
\begin{align}
\underset{
    \bm{\theta}_{\text{comp}}^{s},\, 
    \bm{\theta}_{\text{rec}}^{s},\, 
    \bm{\theta}_{\text{rate}}
}{\min}  
\frac{1}{T} \sum_{t=1}^{T} \sum_{s=1}^{S} \mathcal{L}_{1,t,s}^{\text{LQR}}, \end{align}
where $\mathcal{L}_{1,t,s}^{\text{LQR}} = \left\|
    \bm{y}_t^s - \hat{\bm{y}}_t^s \right\|$, $\{R_t^s\}_{s=1}^S = \mathcal{F}_{\text{rate}}\!\big(\bm{U}_{1,t}; \bm{\theta}_{\text{rate}})$ and $\hat{\bm{y}}_t^s =
    \mathcal{F}_{\text{rec}}^s\!\left(
        \mathcal{Q}\!\left(
            \mathcal{F}_{\text{comp}}^s(\bm{y}_t^s,N_z^s; \bm{\theta}_{\text{comp}}^s),\, R_t^s
        \right); 
        \bm{\theta}_{\text{rec}}^s
    \right) $.

\item \textbf{L2 (Sensing Level):} 
The objective at this level is to minimize the state estimation error. 
The optimization problem is formulated as
\begin{align}
\underset{
    \bm{\theta}_{\text{comp}}^{s},\, 
    \bm{\theta}_{\text{rec}}^{s},\, 
    \bm{\theta}_{\text{rate}}
}{\min}  
\frac{1}{T} \sum_{t=1}^{T} \mathcal{L}_{2,t}^{\text{LQR}},  \end{align}
where  $\mathcal{L}_{2,t}^{\text{LQR}}  = \left\|
    \bm{x}_t^s - \hat{\bm{x}}_t^s \right\|$, $\{R_t^s\}_{s=1}^S = \mathcal{F}_{\text{rate}}\!\big(\bm{U}_{2,t}; \bm{\theta}_{\text{rate}})$ and $\bm{x}_t =
    \mathcal{F}_{\text{KF}}\!\left(
            \mathcal{F}_{\text{rec}}^s\!\left( \mathcal{Q}\!\left(
                    \mathcal{F}_{\text{comp}}^s(\bm{y}_t^s,N_z^s; \bm{\theta}_{\text{comp}}^s),\, R_t^s
                \right);
                \bm{\theta}_{\text{rec}}^s
            \right)
    \right) $.

\item \textbf{L3 (Control Level):} 
The objective is to minimize the long-term \ac{lqr} cost of the closed-loop system, 
reflecting how semantic compression and rate adaptation influence overall control performance. 
The optimization problem is formulated as 
\begin{align}
\underset{
    \bm{\theta}_{\text{comp}}^{s},\, 
    \bm{\theta}_{\text{rec}}^{s},\, 
    \bm{\theta}_{\text{rate}}
}{\min}  
\frac{1}{T} \sum_{t=1}^{T} \mathcal{L}_{3,t}^{\text{LQR}}, \end{align} where $\{R_t^s\}_{s=1}^S = \mathcal{F}_{\text{rate}}\!\big(\bm{U}_{3,t}; \bm{\theta}_{\text{rate}})$ and $\mathcal{L}_{3,t}^{\text{LQR}}  
 = 
\Big(
    (\tilde{\bm{x}}_{t}^{\text{ctl}})^{\top} 
    \bm{Q}_{\mathrm{goal}} 
    \tilde{\bm{x}}_{t}^{\text{ctl}}
    + 
    (\bm{u}_t^{\text{ctl}})^{\top} 
    \bm{R}_{\mathrm{goal}} 
    \bm{u}_t^{\text{ctl}}
\Big).$
\end{itemize}

\color{black}
Following the unified semantic-aware framework presented in Sec.~\ref{sec:Unified-Framework}, 
we propose a semantic-aware communication framework for closed-loop distributed \ac{lqr} systems, 
comprising two training phases as shown in Fig.~\ref{fig:LQRframework}.
\begin{enumerate}
    \item \textbf{Compression Module Training Phase:} 
    During this phase, the semantic compression module is trained using data from a single sensor, corresponding to independently learning an encoder–decoder pair for each sensor and thereby decoupling the compression learning across sensors.
    Such an approach is reasonable in distributed closed-loop \ac{lqr} systems, 
    since the Kalman filter can linearly fuse multi-sensor information, 
    thereby reducing the system’s overall design complexity.
    
    \item \textbf{Rate Adaptation Module Training Phase:} 
    After training the semantic compression module and fixing its parameters, we train the rate adaptation model in a multi-sensor setting. Based on the rate decisions provided by the rate adaptation module, each sensor’s observation is processed through the semantic encoder–decoder pair and subsequently fused in the Kalman filter. The rate controller leverages level-specific state information to optimize the overall closed-loop performance under a given total communication constraint.
\end{enumerate}
Below, we introduce the training process of in detail.

\subsubsection{Compression Module Training Phase}
Building upon the three semantic levels introduced in Sec.~\ref{sec:error-levels}, 
we design a semantic compression algorithm tailored to the \ac{lqr} problem.  
For L1 and L2, training data are collected from trajectories of an uncompressed system and used for offline training of the semantic encoder–decoder pairs. 
For L3, the encoder and decoder are embedded directly into the closed-loop system, where the differentiability of all components allows gradients to be backpropagated through the entire \ac{scc} chain. The detailed training procedures for the three semantic levels are described as follows.

\textbf{L1 Semantic Compression:}
For L1, we consider a single sensor~$s$ and collect $H$ observation trajectories through repeated simulations. 
Let $\mathcal{H} = \{1, 2, \ldots, H\}$ denote the set of trajectory indices, 
and each trajectory $\mathcal{T}_{\text{L1},h}$ consists of $T$ consecutive sensor observations 
$\{\bm{y}_{t}^s, \bm{W}_{t}^{s}\}_{t=1}^T$ with same observation matrix $\bm{C}^s$, where $\bm{y}_{t}^s$ serves as both the input and reconstruction target for training, 
and $\bm{W}_{t}^{s}$ denotes the prior observation covariance matrix that helps improve model performance. 
The complete observation dataset is denoted as 
$\mathcal{D}_{\text{L1}} = \{ \mathcal{T}_{\text{L1},h} \}_{h=1}^{H}$.
During training, the encoder–decoder pair is optimized to minimize the average loss:
$
    \frac{1}{H T}  
    \sum_{\mathcal{T}_{\text{L1},h} \in \mathcal{D}_{\text{L1}}}
    \sum_{t=1}^{T}
    \mathcal{L}_{1,t,s}^{\text{LQR}} ,
$
For each training step, the compression bits $R_t^s$ is randomly sampled 
from a predefined discrete rate set $\mathcal{R}$ 
to enhance robustness under varying communication budgets.

\textbf{L2 Semantic Compression:}
For L2, the training process is extended to incorporate both the true system states and the corresponding sensor observations, 
allowing the encoder–decoder pair to learn semantic representations that are optimal for state estimation. 
Similarly, we consider a single sensor~$s$ and collect $H$ simulation trajectories.  
Each trajectory $\mathcal{T}_{\text{L2},h}$ is defined as a sequence of tuples 
$\{(\bm{x}_{t}, \bm{y}_{t}^{s}, \bm{W}_{t}^{s}, \hat{\bm{x}}_{t-1}, \bm{\Sigma}_{t-1|t-1},\bm{C}^{s})\}_{t=1}^T$, where $\bm{x}_{t}$ serves as the supervision label, 
$\bm{y}_{t}^{s}$ is the input to the semantic compression encoder, and $(\hat{\bm{x}}_{t-1}, \bm{\Sigma}_{t-1|t-1}, \bm{C}^{s})$ are the auxiliary parameters 
that allows the Kalman filter to be incorporated into the training process. 
The complete dataset is denoted by 
$\mathcal{D}_{\text{L2}} = \{ \mathcal{T}_{\text{L2},h} \}_{h=1}^{H}$.
During training, given the randomly sampled compression bits $R_t^s$, the loss can be expressed as
\vspace{-2mm}
\begin{align}
    \mathcal{L}_{\text{train}}^{\text{L2}} 
    = \frac{1}{H T}  
    \sum_{\mathcal{T}_{\text{L2},h} \in \mathcal{D}_{\text{L2}}}
    \sum_{t=1}^{T}
    \mathcal{L}_{2,t}^{\text{LQR}} ,
\end{align}

\textbf{L3 Semantic Compression:}
Unlike L1/L2, 
L3 adopts an online, end-to-end training method. 
Specifically, the encoder and decoder are embedded into the \ac{lqr} system, 
and their parameters are optimized by backpropagating the control-level loss over full rollouts. 
At each training step, we simulate $H$ rollouts with a horizon of $T$ time steps. 
Observations $\bm{y}_t^s = \bm{C}^s \bm{x}_t + \bm{v}_t^s$ are encoded, 
quantized with compression bits $R_t^s$, decoded to $\hat{\bm{y}}_t^s$, 
and fused by the Kalman filter to produce the estimated system state $\hat{\bm{x}}_{t}$. 
The controllable part of the state, $\hat{\bm{x}}_{t}^{\text{ctl}}$, 
is acted upon by the \ac{lqr} law 
$\bm{u}_t^{\text{ctl}} = -\bm{K}_{\text{LQR}}\hat{\bm{x}}_{t}^{\text{ctl}}$. 
The overall control input $\bm{u}_t$ is then formed by combining the remote control command 
$\bm{u}_t^{\text{ctl}}$ and the locally generated command of the uncontrollable target 
$\bm{u}_t^{\text{unctl}}$. 
The system state then evolves according to the linear dynamics $\bm{x}_{t+1} = \bm{A}\bm{x}_t + \bm{B}\bm{u}_t + \bm{w}_t$
after which the next step proceeds iteratively. 
The L3 loss for a rollout is the average quadratic control cost, 
and the overall training minimizes its mean over all rollouts:
\begin{align}
\mathcal{L}_{\text{train}}^{\text{L3}}
= \frac{1}{H T} \sum_{h=1}^{H}\sum_{t=1}^{T}
\Big(
\tilde{\bm{x}}_{t,h}^{\top}\bm{Q}_{\mathrm{goal}}\tilde{\bm{x}}_{t,h}
+ (\bm{u}_{t,h}^{\text{ctl}})^{\top}\bm{R}_{\mathrm{goal}}\bm{u}_{t,h}^{\text{ctl}}
\Big), \nonumber
\end{align}
The entire framework is optimized end-to-end via backpropagation through time, encompassing the Kalman filter prediction, \ac{lqr} feedback, and system dynamics.
The \ac{lqr} gain $\bm{K}_{\text{LQR}}$ is computed based on DARE~\cite{shaiju2008formulas}. 
It is worth noting that although the semantic encoder and decoder are embedded 
into the closed-loop system for training, 
their parameters are shared across all time steps, 
thus avoiding any increase in the model size with longer rollout horizons.

At each semantic level (L1, L2, L3), the encoder–decoder networks share the same architecture (Fig.~\ref{fig:LQRframework}) but are trained with different objectives corresponding to their semantic loss functions.
Leveraging the temporal correlations inherent in closed-loop \ac{scc} systems, we employ a GRU-based \ac{ae} (GRU-AE) to enhance compression efficiency. In the semantic encoder, there are the following four components:
\begin{itemize}
    \item \textbf{MLP:} The raw observation $\bm{y}_t^s$ from sensor $s$ is concatenated with the diagonal elements of the observation noise covariance $\text{diag}(\bm{W}_t^s)$ and the previously transmitted representation $\bm{z}_{t-1}^s$. The diagonal entries indicate the per-dimension uncertainty of the observation, while $\bm{z}_{t-1}^s$ provides the encoder with awareness of the decoder’s prior input. The combined vector is fed into an MLP with input dimension $2N_y^s + N_z^s$, consisting of three layers with 512, 2048, and 512 neurons, respectively.

    \item \textbf{GRU:} The MLP output, together with the previous historical embedding $\bm{o}_{t-1}^{s,\text{enc}}$, is fed into a GRU, which updates the embedding to $\bm{o}_t^{s,\text{enc}}$ and produces the intermediate representation for the current time step. The GRU embedding dimension is fixed at 512.
    
    \item \textbf{MLP:} An MLP compresses the intermediate representation to a fixed dimension $N_z^s$, generating $\tilde{\bm{z}}_{t}^{s}$.

     \item \textbf{Quantization:} $\tilde{\bm{z}}_{t}^{s}$ is quantized by the quantization layer according to the given compression bits $R_t^s$ to obtain the semantic representation $\bm{z}_{t}^{s}$ for transmission. 
    
\end{itemize}

The decoder mirrors the encoder and operates as follows:
\begin{itemize}
    \item \textbf{MLP:} The received semantic representation $\bm{z}_{t}^{s}$, the observation noise level $\text{diag}(\bm{W}^s)$, and the corresponding compression bits $R_t^s$ are concatenated and fed into an MLP with input and output dimensions of $N_z^s + N_y^s + 1$ and 512, respectively.

    \item \textbf{GRU layer:} The MLP output, together with the previous historical embedding $\bm{o}_{t-1}^{s,\text{dec}}$, is fed into a GRU that updates the embedding to $\bm{o}_{t}^{s,\text{dec}}$ and generates an intermediate representation $\bm{z}_{t}^{s,\text{dec}}$. 
    The GRU embedding dimension is fixed at 512.

    \item \textbf{MLP:} The intermediate representation is fed into another MLP that outputs both the reconstructed observation $\hat{\bm{y}}_t^s$ and the uncertainty estimate $\bm{\zeta}_t^s$ (where $\bm{\zeta}_t^s = \text{diag}(\hat{\bm{W}}_t^s)$ in the \ac{lqr} case). The uncertainty output, used in L2 and L3 training, enables the state tracker to adapt its estimation to compression-induced distortions. The MLP has an input dimension of 512 and three layers with 2048, 512, and $N_y^s$ neurons, respectively.
\end{itemize}

\subsubsection{Rate Adaptation Algorithm}
Given a trained semantic compression module, we design a dynamic rate adaptation mechanism for each semantic level $l \in \{1, 2, 3\}$. 
Similar to problem~\eqref{Prob:RateAdapt}, the objective is to minimize the long-term semantic loss subject to the total rate constraint:
\begin{align}
\min_{\bm{\theta}_{\text{rate}}} \quad 
& \frac{1}{T} \sum_{t=1}^{T} 
    \mathcal{L}_{l,t}^{\text{LQR}}  , \nonumber \\
\text{s.t.} \quad 
& \{R_t^s\}_{s=1}^S  = \mathcal{F}_{\text{rate}}(\bm{U}_{l,t}; \bm{\theta}_{\text{rate}}), \\
&  \frac{1}{T} \sum_{t=1}^{T} 
    \sum_{s=1}^{S} R_t^s \le R,
    \label{equ:longterm_rate_constraint}
\end{align}
where $\mathcal{L}_{l,t}^{\text{LQR}} $ is the LQR-based semantic loss at level $l$, defined in Sec. \ref{subsec:lqrImplementation},
$\bm{U}_{l,t}$ represents the level-specific prior information, which will be detailed later in this subsection, and $\bm{\theta}_{\text{rate}}$ are the trainable parameters of the policy network. 
The long-term constraint in~\eqref{equ:longterm_rate_constraint} ensures that the average transmission cost does not exceed the total system bandwidth~$R$.

To satisfy the long-term rate constraint, 
we employ a Lyapunov optimization framework by introducing a virtual queue $V_t$ that tracks the accumulated rate surplus~\cite{sun2025energy}: $V_{t+1} = V_t + \sum_{s=1}^S R_t^s - R$.
If the virtual queue $V_t$ is stable, the long-term rate constraint in~\eqref{equ:longterm_rate_constraint} is satisfied.
We define the quadratic Lyapunov function $H(V_t) = \tfrac{1}{2}V_t^2$, 
whose one-step drift satisfies $H(V_{t+1}) 
    \le \tfrac{1}{2}V_t^2 + B 
        + V_t (\sum_{s=1}^S R_t^s - R)$
where $B = \tfrac{1}{2}\!(\sum_{s=1}^S R_t^s)^2 + \tfrac{1}{2}R^2$ is a finite constant upper bound. 
The Lyapunov drift is thus
$  \Delta(V_t) 
    = H(V_{t+1}) - H(V_t)
    \le B + V_t (\sum_{s=1}^S R_t^s - R)$.
According to ~\cite{sun2025energy}, minimizing the drift at each time slot is an effective way to stabilize the queue. To simultaneously ensure queue stability and optimize performance, we minimize a weighted combination of drift and the instantaneous semantic loss, leading to the single-slot optimization problem:
\begin{align}
    \min_{\bm{\theta}_{\text{rate}}} \quad & \mathcal{L}_{l,t}^{\text{LQR}}  + \beta V_t \sum_{s=1}^S R_t^s, \\
    \text{s.t.} \quad & \{R_t^s\}_{s=1}^S = \mathcal{F}_{\text{rate}}(\bm{U}_{l,t}; \bm{\theta}_{\text{rate}}),
\end{align}
where $\beta>0$ controls the trade-off between estimation accuracy and long-term rate constraint satisfaction. As $V_t$ grows, the policy is increasingly penalized for allocating excessive bandwidth, thereby promoting long-term feasibility.

We further formulate rate adaptation as a Markov decision process (MDP) 
and employ the \ac{ppo}~\cite{pan2024quality} to learn the rate policy $\mathcal{F}_{\text{rate}}(\cdot)$. 
The agent selects rate configurations for each time step to minimize the level-specific semantic loss while respecting the rate constraint. The \ac{ppo}-based rate adaptation agent is defined as follows:
\begin{itemize}
    \item \textit{State:} The semantic-level prior $\bm{U}_{l,t}$ represents the prior information for different level considerations. In \ac{lqr} case, we set $\bm{U}_{1,t} = \{\bm{W}_t^{1:S}, V_t\}$ for L1, and $\bm{U}_{2,t} = \{\bm{W}_t^{1:S}, V_t,\hat{\bm{\Sigma}}_{t-1|t-1}, \hat{\bm{x}}_{t-1}\}$ for L2, and $\bm{U}_{3,t} = \{\bm{W}_t^{1:S}, V_t,\hat{\bm{\Sigma}}_{t-1|t-1}, \hat{\bm{x}}_{t-1},\bm{x}^{\text{desired}}_{t}\}$ for L3.

    \item \textit{Action:} The action at time~$t$ is the rate configuration 
    $a_t = [R_t^1, \dots, R_t^S]$, 
    where each $R_t^s$ is chosen from a discrete rate set 
    $ \mathcal{R} = \{0, N_z^s, 2N_z^s, \dots, 6N_z^s\}$. 0 indicates that no transmission is performed for sensor~$s$.

    \item \textit{Reward:} The level-$l$ reward at time $t$ is given by: $        r_{l,t}^{\text{reward}} =\mathcal{L}_{l,t}^{\text{LQR}}  + \beta V_t \sum_{s=1}^{S} R_t^s$,
    where a higher reward reflects reduced semantic loss or \ac{lqr} cost and tighter satisfaction of the rate constraint, balancing performance and communication efficiency. It is worth noting that when 0 bits are assigned in a given time slot, the reconstructed observation is set to 0 for error computation, which indicates that even without transmission, the L1 scheme still incurs reconstruction loss.
\end{itemize}
To optimize the rate adaptation policy, we employ the GRU-based \ac{ppo} algorithm, a policy-gradient method that iteratively updates the policy network by maximizing a clipped surrogate objective. During training, three networks are maintained: 1) the actor network parameterized by $\bm{\theta}_{\text{rate}}$ which generates the rate allocation policy based on the semantic-level state sequence; 2) the old actor network $\bm{\theta}_{\text{rate}}^{\text{old}}$ with parameters, which stores the policy from the previous iteration and serves as the baseline for computing probability ratios; 3) the critic network parameterized by $\bm{\eta}$, which estimates the state value function $O_{\bm{\eta}}(\bm{U}_{l,t})$. Let $\pi_{\bm{\theta}_{\text{rate}}}(a_t \! \mid \! \bm{U}_{l,t})$ denote the stochastic policy realized by the GRU-based actor network, which outputs the rate allocation action $a_t$.
The actor loss function is defined as
\begin{align}
    L(\bm{\theta}_{\text{rate}})  =  -
    \frac{1}{T}\sum_{t=1}^{T}  \! &\left[
        \min \!\left(
            \rho_t(\bm{\theta}_{\text{rate}}) \, \hat{A}_t,\, \right.\right. \nonumber\\
        &\ \left.\left.  \text{clip}\big(\rho_t(\bm{\theta}_{\text{rate}}), 1-\epsilon, 1+\epsilon\big)\, \hat{A}_t
        \right)
    \right],
\end{align}
where $\rho_t(\bm{\theta}_{\text{rate}}) 
= \tfrac{\pi_{\bm{\theta}_{\text{rate}}}(a_t \mid \bm{U}_{l,t})}
        {\pi_{\bm{\theta}_{\text{rate}}^{\text{old}}}(a_t \mid \bm{U}_{l,t})}$ 
is the probability ratio between the new and old policies,
and $\epsilon$ is a clipping threshold that stabilizes training. 
The advantage estimate $\hat{A}_t$ is computed using the generalized advantage estimation (GAE) method. 
We first evaluate the temporal-difference (TD) residual
$\delta_t 
    = r_{l,t}^{\mathrm{reward}}
    + \gamma\, O_{\bm{\eta}}\!\left(\bm{U}_{l,t+1}\right)
    - O_{\bm{\eta}}\!\left(\bm{U}_{l,t}\right)$,
and then accumulate advantages backward along each trajectory via 
$ \hat{A}_t 
    = \delta_t 
    + \gamma \lambda\, \hat{A}_{t+1}$,
where $\gamma \in (0,1)$ is the discount factor and $\lambda \in (0,1)$ 
is the GAE smoothing parameter. 
The critic network is trained by regressing its value predictions toward the discounted return computed along each trajectory:
\begin{align}
    L_{\text{critic}}(\bm{\eta})
    =
   \frac{1}{T}\sum_{t=1}^{T} \Big[
        \big(
            O_{\bm{\eta}}(\bm{U}_{l,t})
            -
            \hat{O}_t^{\text{return}}
        \big)^2
    \Big],
\end{align}
where
$\hat{O}_t^{\text{return}}
    = \sum_{t'=t}^{T}
        \gamma^{\,t'-t}\,
        r_{l,t'}^{\mathrm{reward}}$. 
During training, the parameters of the old actor network are periodically updated to the latest trained weights $\bm{\theta}_{\text{rate}}$ every $E_{\text{update}}$ epochs.

The actor and critic networks share a GRU layer with an output dimension of 256, which captures the temporal dependencies of sequential states. 
Subsequently, the actor network outputs rate allocations with a dimension of $S \times N_z^s$, while the critic network produces a scalar value estimate with 1 dimension. 
Through \ac{ppo} training, the policy network $\mathcal{F}_{\text{rate}}(\cdot)$ learns to allocate rates dynamically across sensors and time steps, adapting to semantic importance, communication conditions, and task requirements. 
This enables closed-loop \ac{scc} systems to achieve goal-aligned performance while satisfying practical bandwidth constraints.

\color{black}

\section{Results}

\subsection{Experimental Setup and Baselines}

\begin{figure*}[t]
    \centering
    \captionsetup[subfigure]{justification=centering, font=footnotesize, skip=5pt}

    \begin{minipage}{0.22\textwidth}
        \centering
        \subfloat[Uncompressed communication\label{fig:traj_a}]{
            \includegraphics[width=\linewidth]{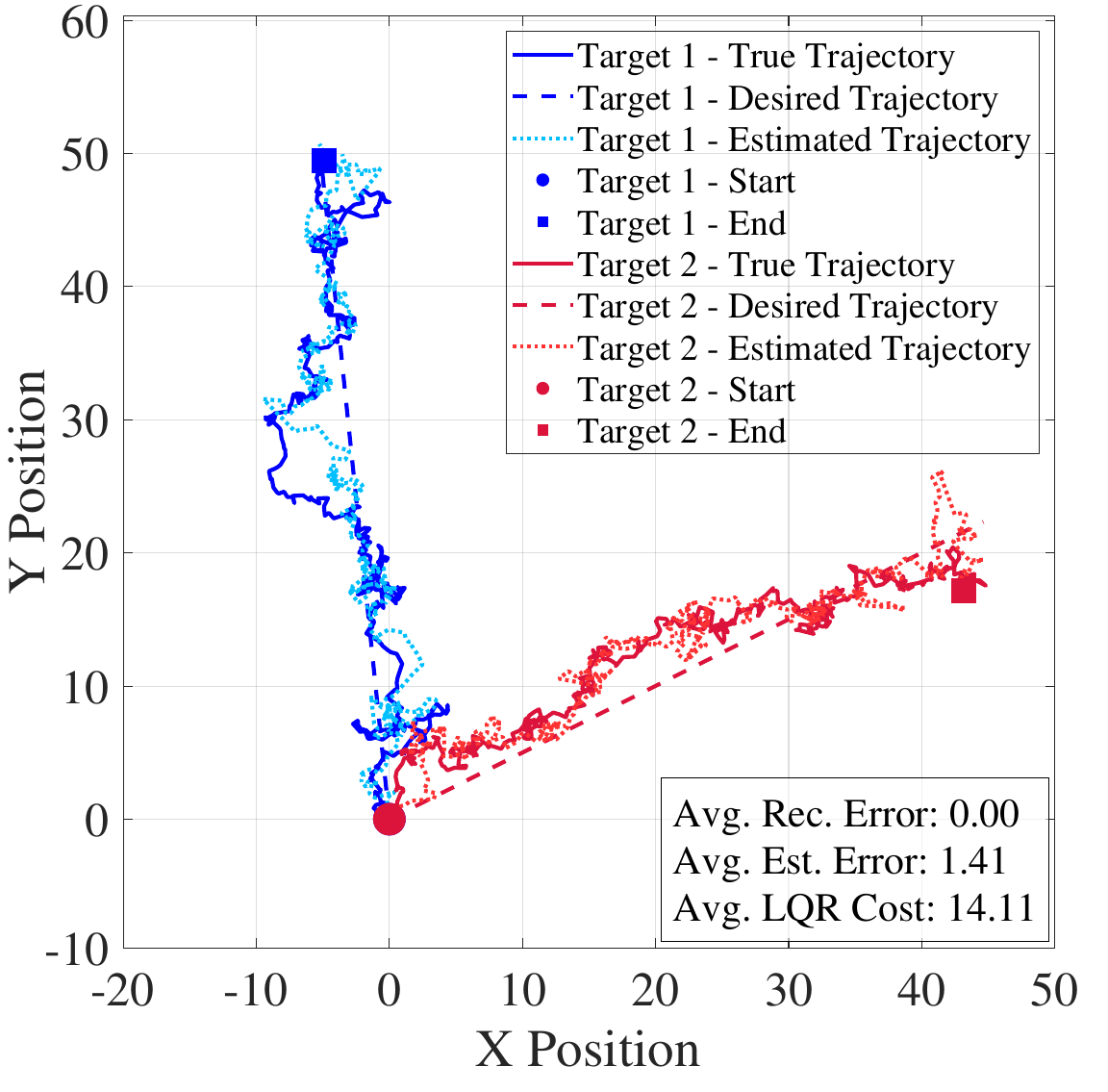}
        }
    \end{minipage}
    \hfill
    \begin{minipage}{0.22\textwidth}
        \centering
        \subfloat[No control commands\label{fig:traj_b}]{
            \includegraphics[width=\linewidth]{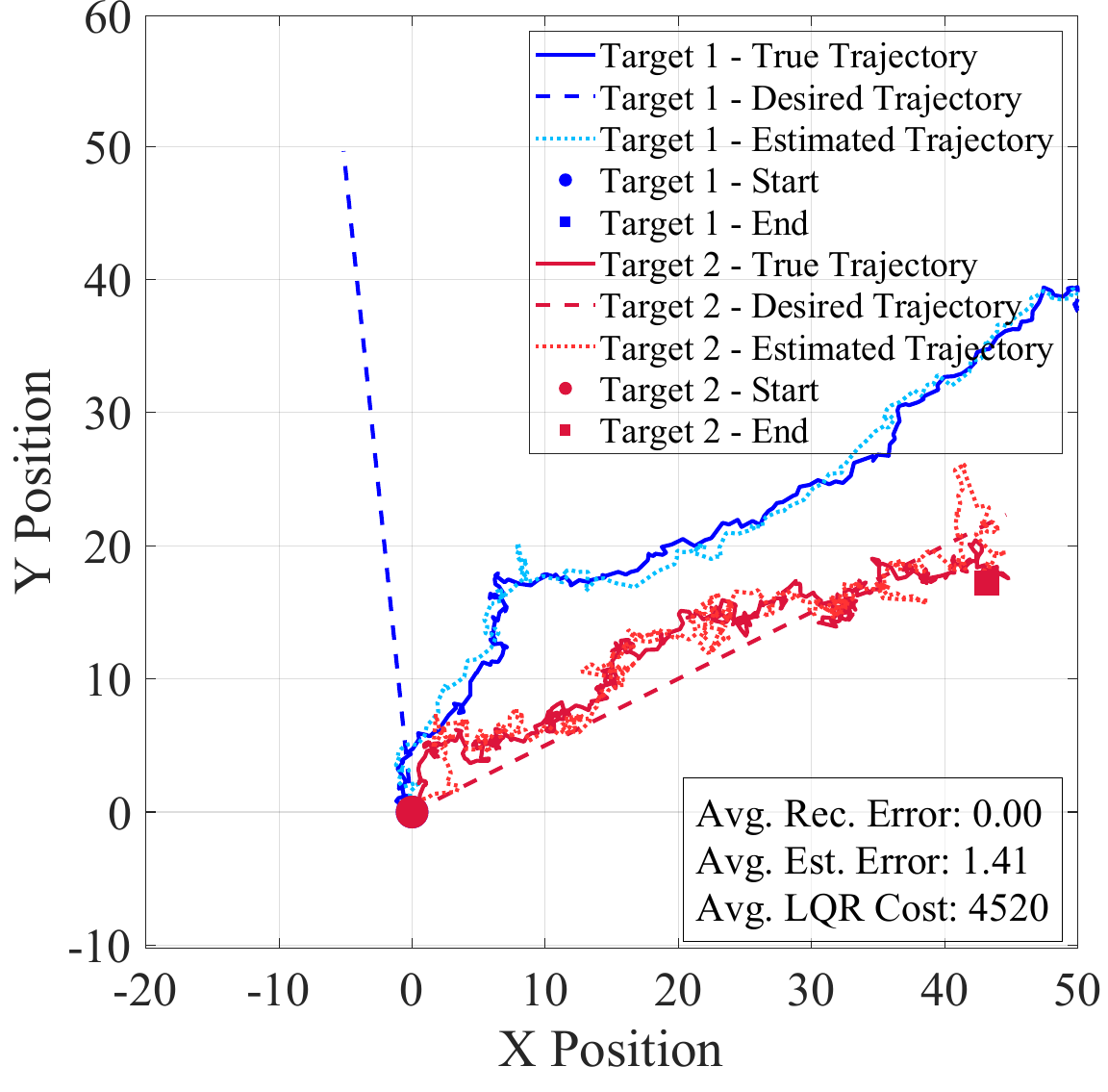}
        }
    \end{minipage}
    \hfill
    \begin{minipage}{0.22\textwidth}
        \centering
        \subfloat[L1 AE  \label{fig:traj_c}]{
            \includegraphics[width=\linewidth]{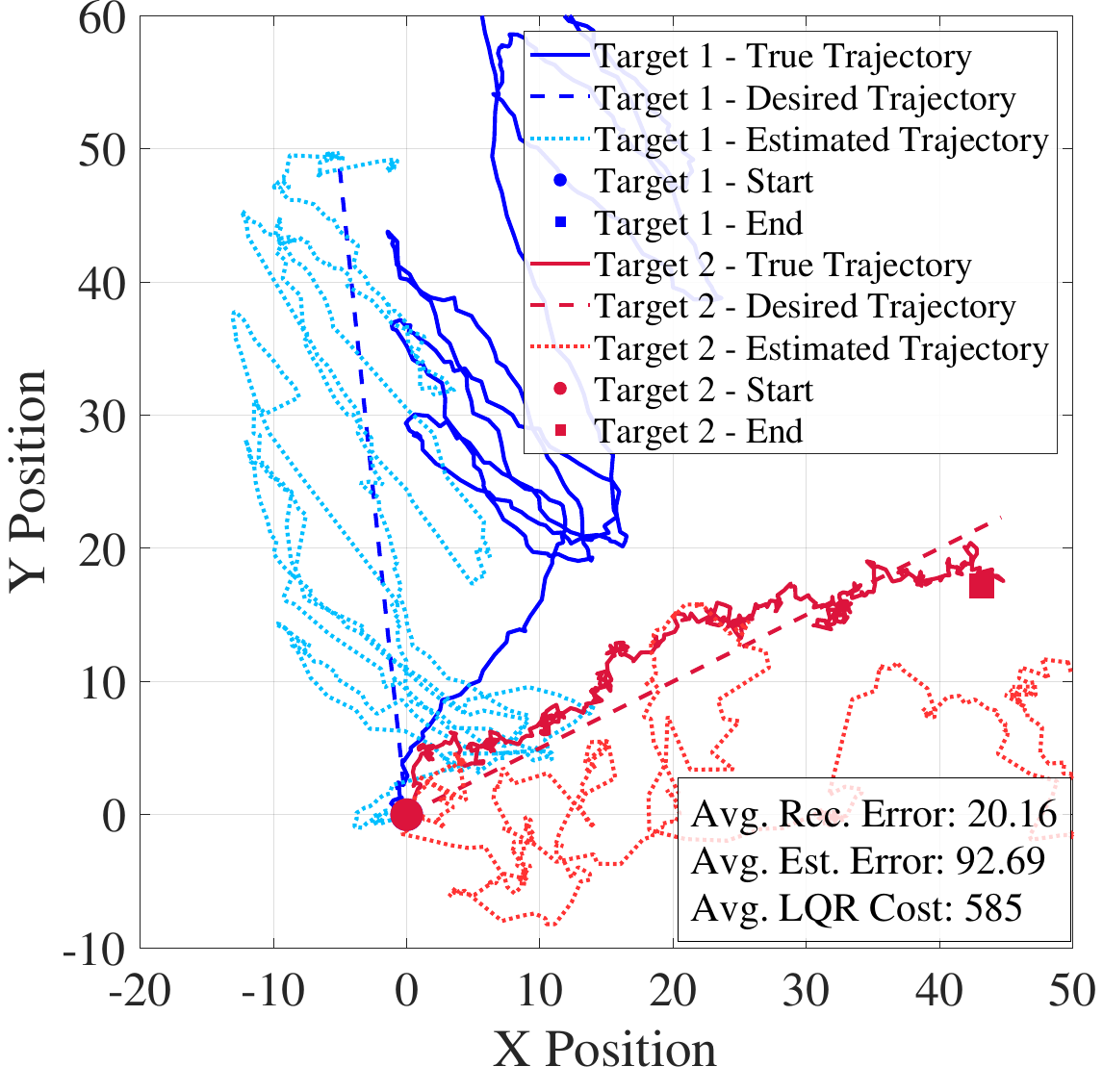}
        }
    \end{minipage}
    \hfill
    \begin{minipage}{0.22\textwidth}
        \centering
        \subfloat[L2 AE \label{fig:traj_c}]{
            \includegraphics[width=\linewidth]{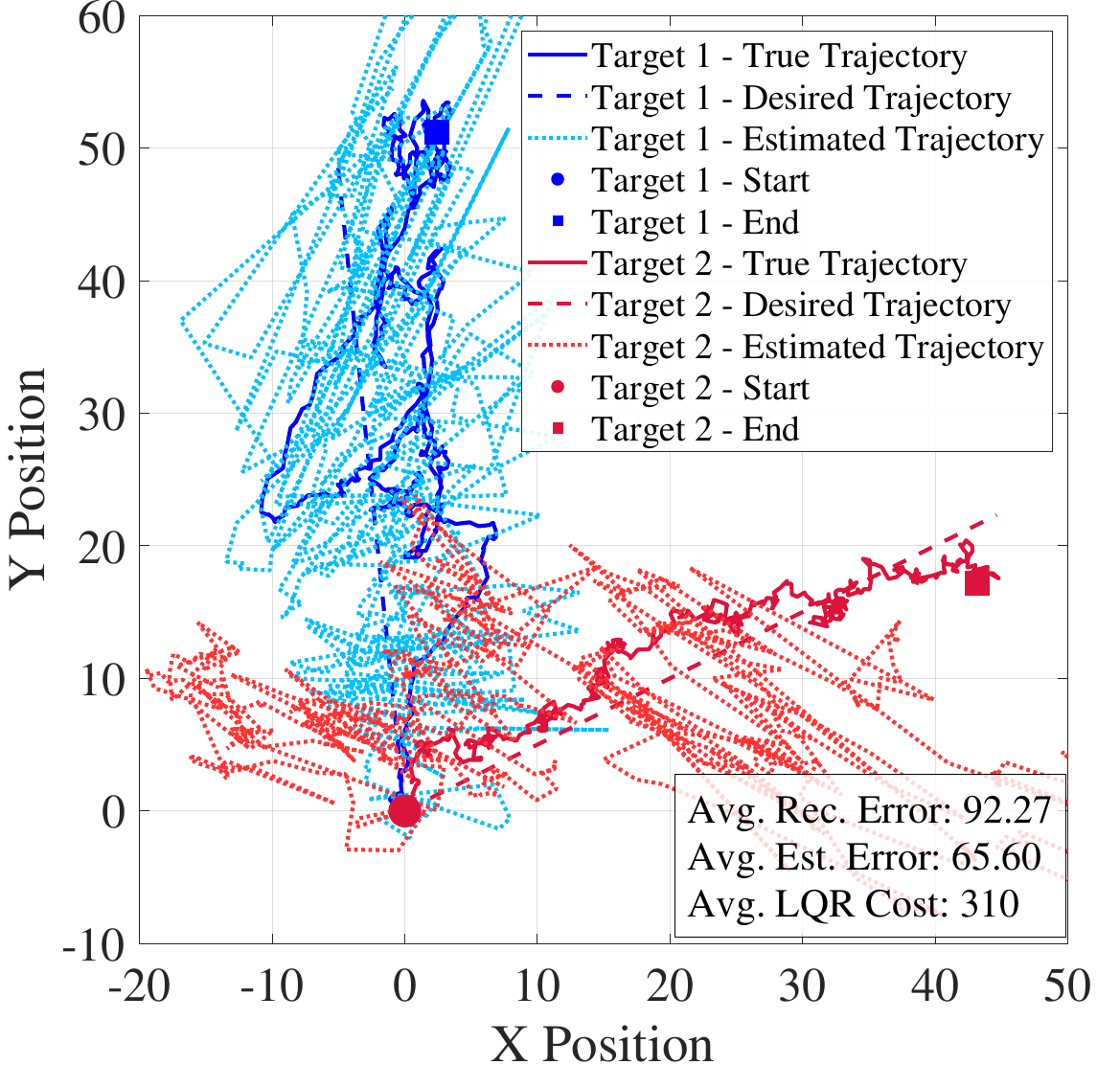}
        }
    \end{minipage}    
    \vspace{-5mm}
    \vskip\baselineskip
    \begin{minipage}{0.22\textwidth}
        \centering
        \subfloat[L3 AE \label{fig:traj_d}]{
            \includegraphics[width=\linewidth]{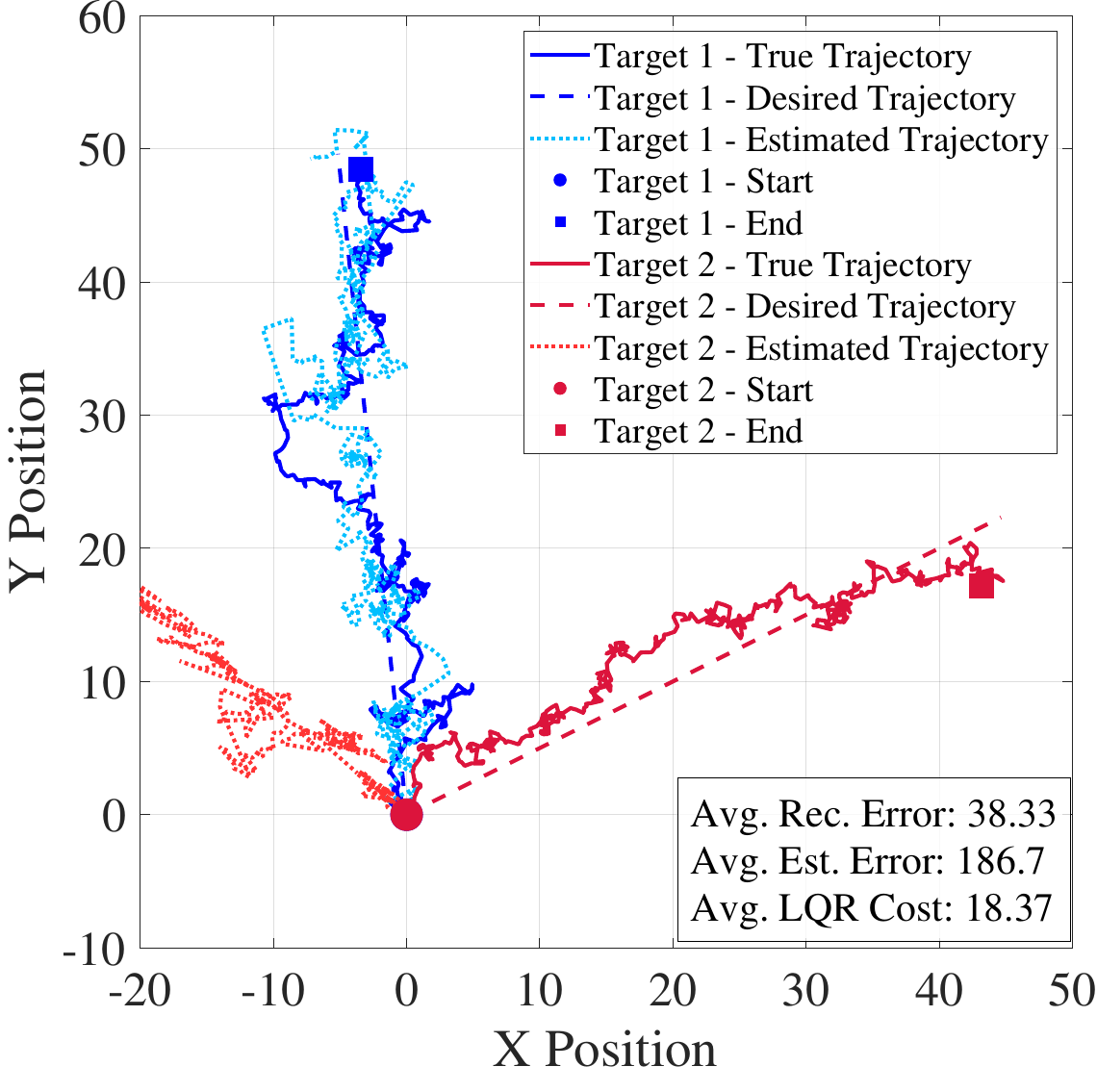}
        }
    \end{minipage}
    \hfill
    \begin{minipage}{0.22\textwidth}
        \centering
        \subfloat[L1 GRU-AE \label{fig:traj_e}]{
            \includegraphics[width=\linewidth]{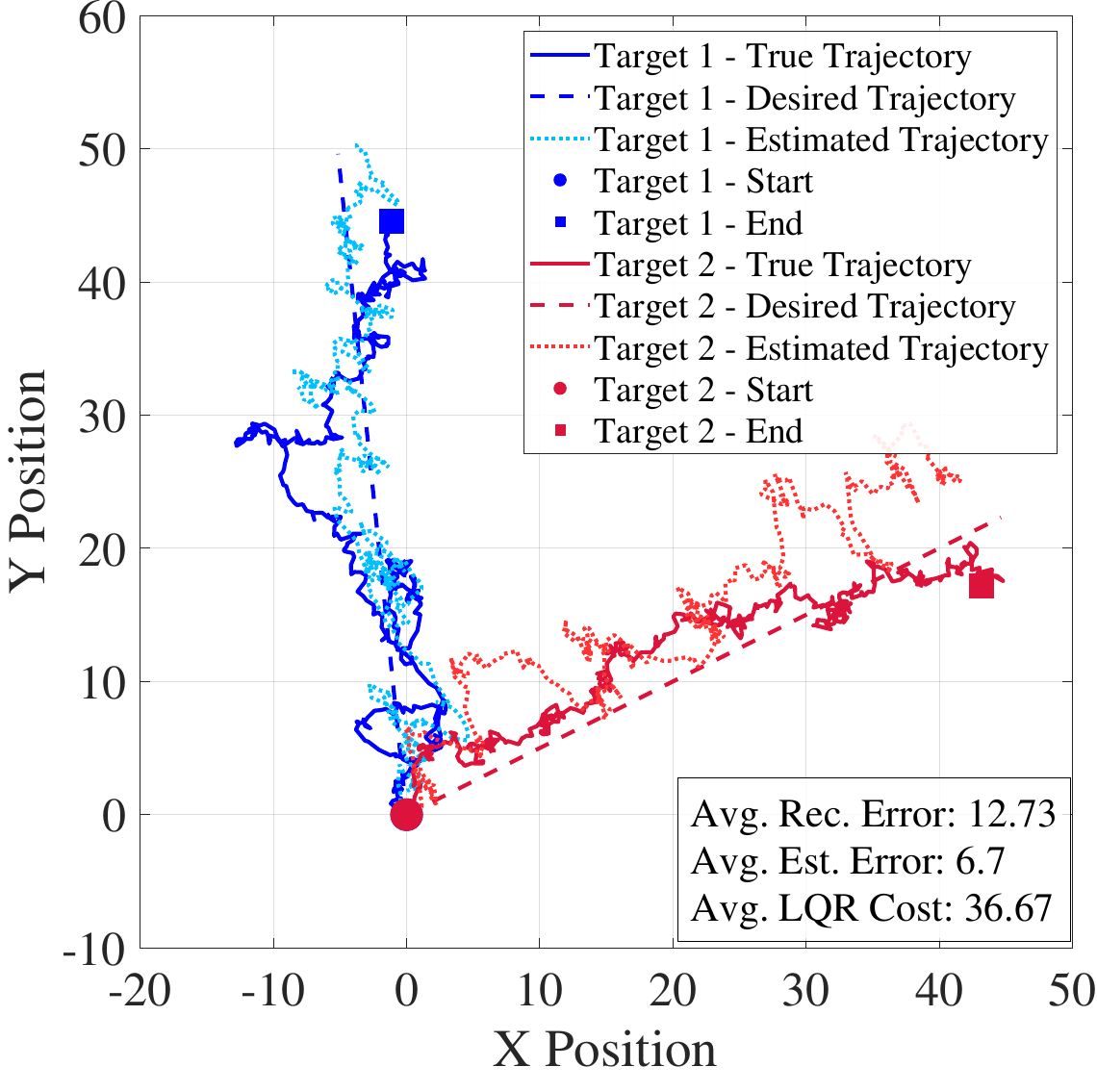}
        }
    \end{minipage}
    \hfill
    \begin{minipage}{0.22\textwidth}
        \centering
        \subfloat[L2 GRU-AE \label{fig:traj_f}]{
            \includegraphics[width=\linewidth]{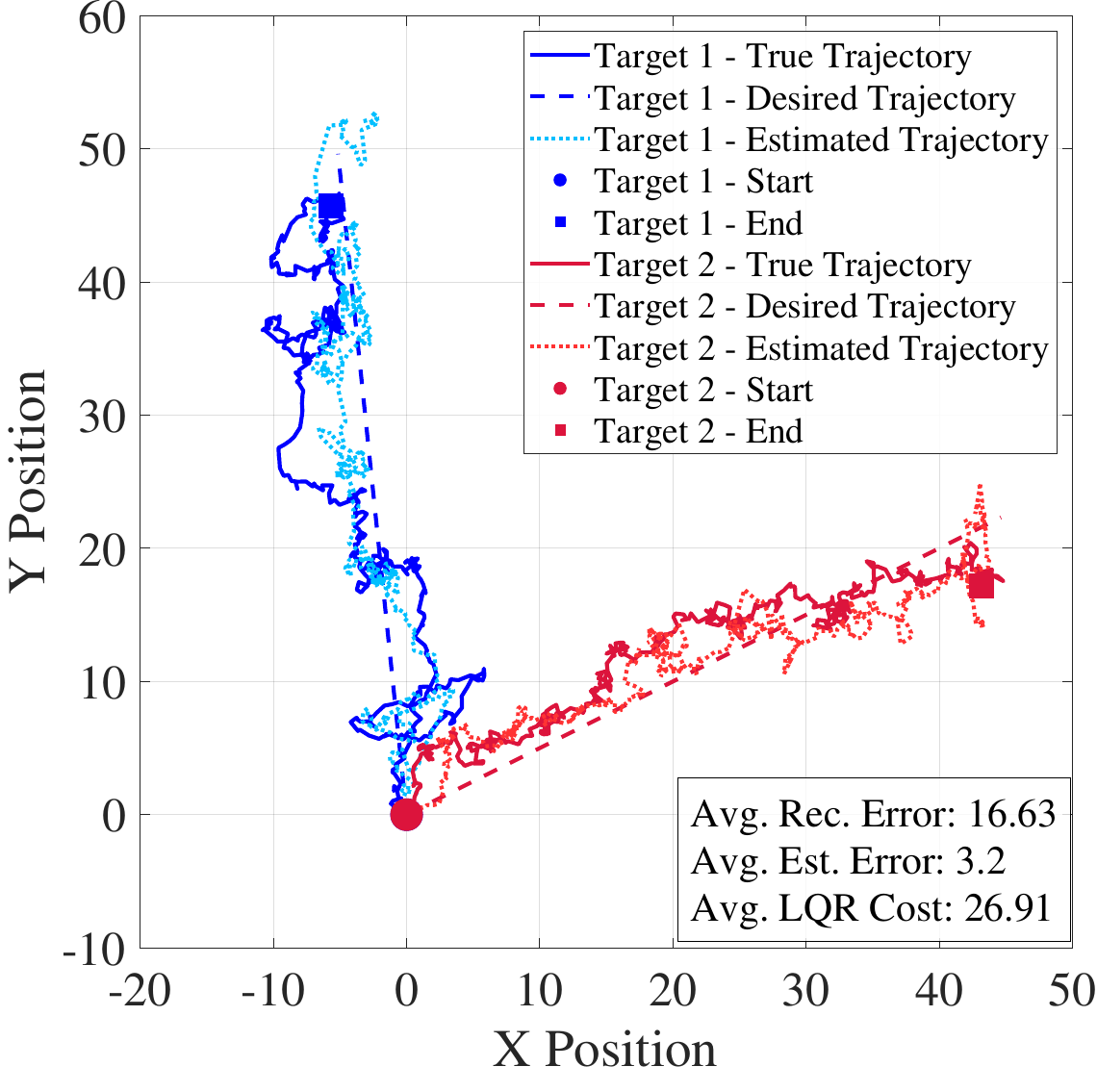}
        }
    \end{minipage}
    \hfill
    \begin{minipage}{0.22\textwidth}
        \centering
        \subfloat[L3 GRU-AE \label{fig:traj_f}]{
            \includegraphics[width=\linewidth]{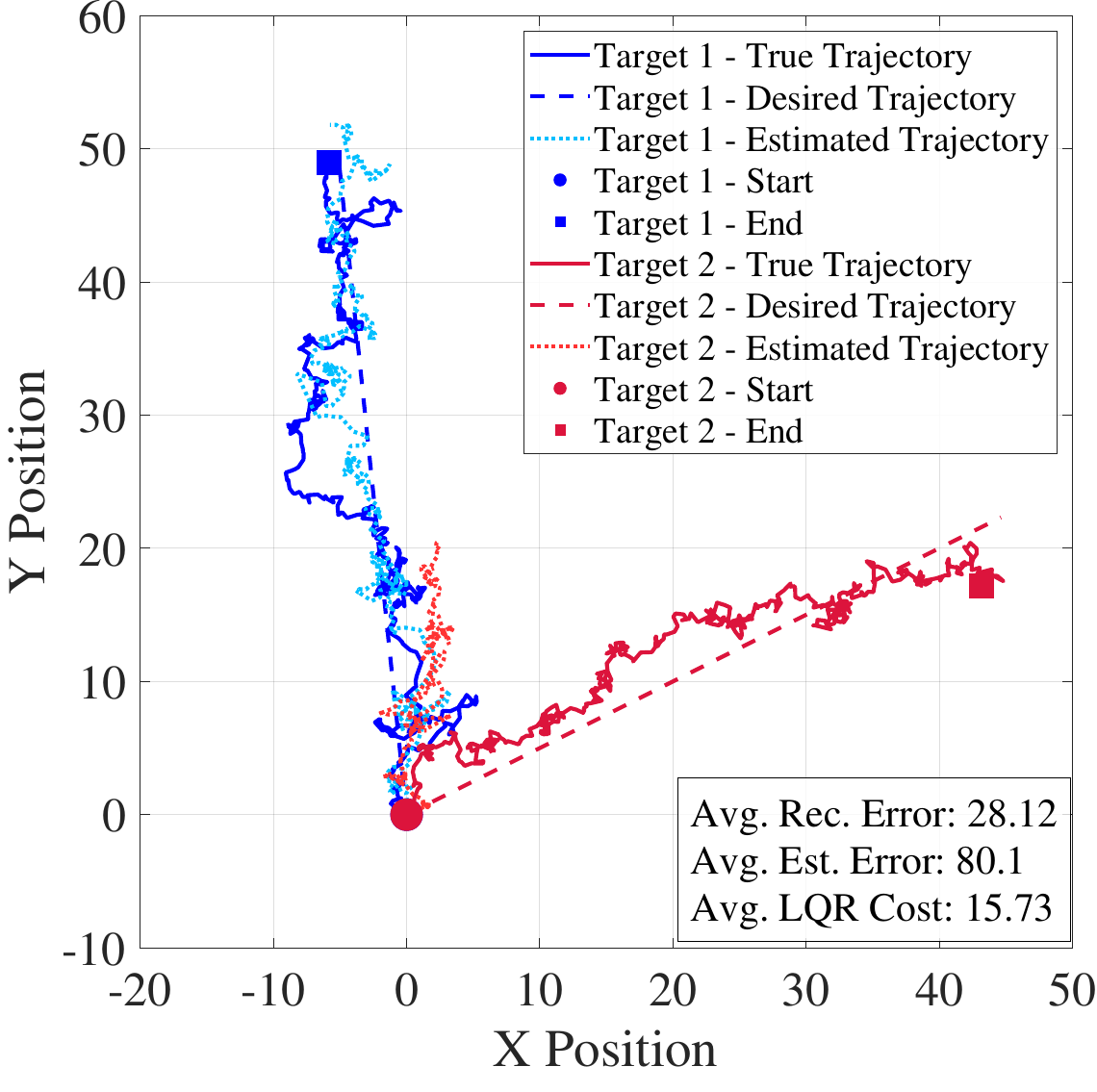}
        }
    \end{minipage}
    \caption{
Trajectory results under different communication and compression settings.
The red and blue curves represent the trajectories of the uncontrollable and controllable targets, respectively. 
Subfigures (a)--(b) show the baseline cases with uncompressed communication and no control commands. 
Subfigures (c)--(h) present the results of different AE-based and GRU-AE–based compression schemes.
}
    \label{fig:traj_compression}
\end{figure*}

\begin{figure*}[t]
    \centering
    \vspace{-3mm}

    \begin{minipage}{0.32\textwidth}
        \centering
        \subfloat[Results for reconstruction error.]{
            \includegraphics[width=\linewidth]{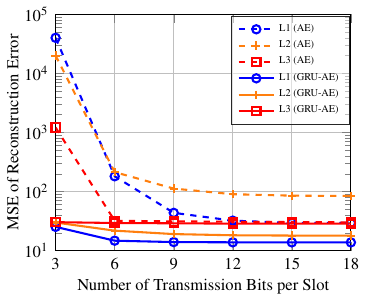}
            \label{fig:result_compression-figure1}
        }
    \end{minipage}
    \hspace{3pt}
    \begin{minipage}{0.32\textwidth}
        \centering
        \subfloat[Results for estimation error.]{
            \includegraphics[width=\linewidth]{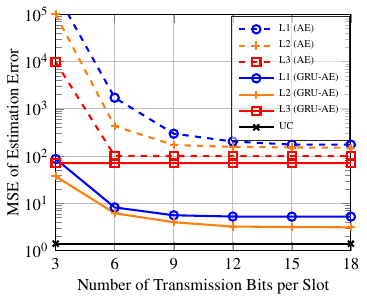}
            \label{fig:result_compression-figure2}
        }
    \end{minipage}
    \hspace{3pt}
    \begin{minipage}{0.32\textwidth}
        \centering
        \subfloat[Results for \ac{lqr} cost.]{
            \includegraphics[width=\linewidth]{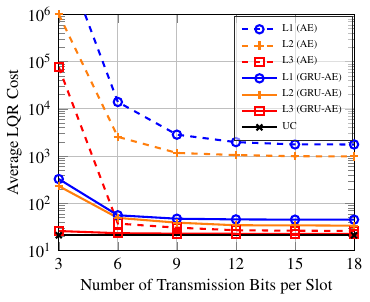}
            \label{fig:result_compression-figure3}
        }
    \end{minipage}

    \caption{Comparison of different semantic compression levels in a single-sensor setup. The total transmission bits per slot range from 3 to 18, corresponding to 1–6 quantization bits under a 3-dimensional compression setting.}
    \label{fig:result_compression}
    \vspace{-3mm}
\end{figure*}


\subsubsection{LQR Parameter}
We consider a vehicle control scenario for evaluating the proposed semantic-aware framework. Both controllable and uncontrollable objects follow a discrete-time linear dynamical model with system matrices
$\bm{A}^{\text{ctl}} = \bm{A}^{\text{unctl}}=
\left[
\begin{smallmatrix}
1 & 0 & \Delta t & 0\\
0 & 1 & 0 & \Delta t\\
0 & 0 & 1 & 0\\
0 & 0 & 0 & 1
\end{smallmatrix}
\right]$ and $ \bm{B}^{\text{ctl}} = \bm{B}^{\text{unctl}}=
\left[
\begin{smallmatrix}
0.5\Delta t^2 & 0 \\
0 & 0.5\Delta t^2 \\
\Delta t & 0 \\
0 & \Delta t
\end{smallmatrix}
\right]$,
where $\Delta t = 0.1$. The state dimension is $N_x^{\text{ctl}} = N_x^{\text{unctl}} = 4$, corresponding to the two-dimensional position and velocity of each object, including one remotely controlled vehicle and one autonomous (uncontrollable) vehicle. The control input dimension is $N_u^{\text{ctl}} = N_u^{\text{unctl}} = 2$, representing the acceleration along the $x$- and $y$-axes. We also set $\hat{\bm{x}}_{0|0} = \bm{x}_0 = [0, 0, v_x^{\text{ctl,init}}, v_y^{\text{ctl,init}}, 0, 0, v_x^{\text{unctl,init}}, v_y^{\text{unctl,init}}]$.
The desired states evolve according to 
$\begin{bmatrix}
        \bm{x}_{t+1}^{\text{ctl,desired}} \\
        \bm{x}_{t+1}^{\text{unctl,desired}}
    \end{bmatrix} = \bm{A} \begin{bmatrix}
        \bm{x}_t^{\text{ctl,desired}} \\
        \bm{x}_t^{\text{unctl,desired}}
    \end{bmatrix}$. 
For the self-controlled (uncontrollable) object, the control input $\bm{u}_t^{\text{unctl}}$ is set to the ideal action computed based on its true state, representing a perfect control command. We also set $\bm{\Sigma}_{0|0} = 0.0001\,\bm{I}$, $\bm{Q} = \bm{I}$, and $\bm{W}^s = \bm{I}$. The \ac{lqr} cost uses $\bm{Q}_{\text{goal}} = \bm{I}$ and $\bm{R}_{\text{goal}} = \bm{I}$. We set $N_z^s =3$ for a fixed compression dimension. Considering the different magnitudes of semantic losses across levels, we set $\beta = 0.05$ for L1, $\beta = 0.01$ for L2, and $\beta = 0.03$ for L3 to balance performance and communication efficiency.

\subsubsection{Dynamic Environment Setup}

We consider a dynamic environment where observation noise vary over time. In terms of observation noise, each sensor can be in one of two states at any time slot $t$: a \textit{good} state or a \textit{bad} state. The observation noise covariance is defined as:
\begin{align}
    \bm{W}_t^s =
\begin{cases}
\bm{I}, & \text{if sensor } s \text{ is in the \textit{good} state} , \\
30\bm{I}, & \text{if sensor } s \text{ is in the \textit{bad} state}.
\end{cases}
\end{align}

\subsection{Results for the Semantic Compression Module}


\begin{figure*}[t]
    \centering
    \captionsetup[subfigure]{justification=centering, font=footnotesize, skip=5pt} 
    \begin{minipage}{0.32\textwidth}        
        \centering
        \subfloat[Results for reconstruction error.]{
            \includegraphics[width=\linewidth]{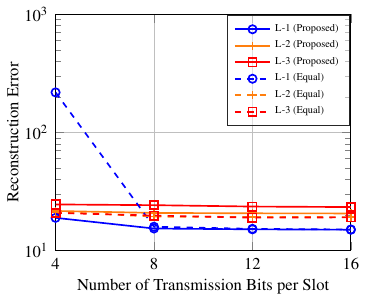}
            \label{fig:result_all_figure1}
        }
    \end{minipage}
    \begin{minipage}{0.32\textwidth}    
        \centering
        \subfloat[Results for estimation error.]{
            \includegraphics[width=\linewidth]{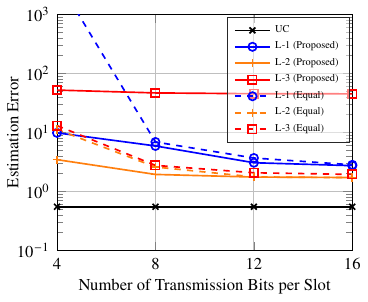}
            \label{fig:result_all_figure2}
        }
    \end{minipage}
    \begin{minipage}{0.32\textwidth}    
        \centering
        \subfloat[Results for \ac{lqr} cost.]{
            \includegraphics[width=\linewidth]{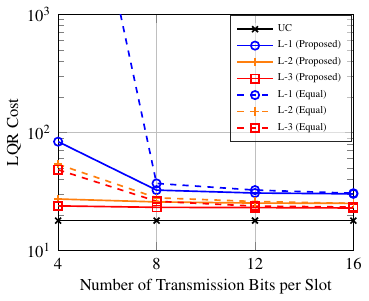}
            \label{fig:result_all_figure3}
        }
    \end{minipage}
    \caption{Performance comparison under different long-term communication budgets.}
    \label{fig:result_all}
    \vspace{-5mm}
\end{figure*}

We first evaluate the performance of different semantic compression algorithms at various semantic levels in a single-sensor configuration. In this experiment, rate adaptation is disabled, and the single sensor transmits with a fixed number of bits per slot. Each element of $\bm{C}^s \in \mathbb{R}^{8 \times 20}$ is independently sampled from a uniform distribution $[-0.2, 0.2]$, indicating that the sensor can simultaneously observe the states of both controllable and uncontrollable objects. The noise state of each sensor evolves as an independent two-state Markov process with the following state transition matrix:
$
\bm{P} =
\begin{bmatrix}
p_{gg} & p_{gb} \\
p_{bg} & p_{bb}
\end{bmatrix},
$
where $p_{gg}=p_{bb} =0.9$ is the probability of remaining in the previous state, $p_{gb} = p_{bg} = 0.1$ denotes the probabilities of switching between the two states. $v_x^{\text{ctl,init}}$ and $v_y^{\text{ctl,init}}$ are randomly generated initial velocities satisfying $v_x^{\text{ctl,init}} \in [-1, 1]$ and $(v_x^{\text{ctl,init}})^2 + (v_y^{\text{ctl,init}})^2 = 1$. 
The initialization of $v_x^{\text{unctl,init}}$ and $v_y^{\text{unctl,init}}$ follows the same procedure. 

Fig.~\ref{fig:traj_compression} illustrates a representative trajectory example under different settings, where $R_t^s=6$ bits are used for quantization.
The red curve represents the trajectory of the uncontrollable (self-controlled) target (target 2), while the blue curve corresponds to the controllable target (target 1). AE refers to the baseline proposed in~\cite{pan2025rate}, and GRU-AE represents the proposed model. 
As expected, the uncontrollable target consistently follows its desired path regardless of communication or control settings.
With uncompressed communication, the controllable target closely follows the desired trajectory, as shown in Fig.~\ref{fig:traj_compression}(a).
Without closed-loop control, the controllable target deviates significantly from the desired path due to the absence of feedback, as shown in Fig.~\ref{fig:traj_compression}(b).
Fig.~\ref{fig:traj_compression}(c) and Fig.~\ref{fig:traj_compression}(d) show that both the L1 AE and L2 AE methods lead to system instability due to excessive compression.
In comparison, the AE-GRU–based L1 and L2 methods maintain stability (as shown in Fig.~\ref{fig:traj_compression}(f) and Fig.~\ref{fig:traj_compression}(g)), benefiting from the GRU’s ability to extract temporal dependencies.
Furthermore, the L3 AE and L3 GRU-AE approaches improve the control performance of target~1 by disregarding the state estimation of target~2, which explains the relatively large estimation errors as observed in Fig.~\ref{fig:traj_compression}(e) and Fig.~\ref{fig:traj_compression}(h).

The results in Fig.~\ref{fig:result_compression} illustrate the impact of semantic compression design on reconstruction, estimation, and control performance for a single-sensor setup under different transmission bit budgets per slot.
“UC” denotes the uncompressed baseline.
Fig.~\ref{fig:result_compression}\subref{fig:result_compression-figure1} reports the reconstruction error, Fig.~\ref{fig:result_compression}\subref{fig:result_compression-figure2} presents the state estimation error, and Fig.~\ref{fig:result_compression}\subref{fig:result_compression-figure3} shows the resulting \ac{lqr} cost. 
The GRU-AE consistently outperforms the AE across all metrics, demonstrating the benefit of exploiting temporal correlations in sequential observations for more effective semantic compression. In terms of reconstruction performance, both AE and GRU-AE achieve lower reconstruction errors as the number of transmitted bits increases. 
Among them, the L1 GRU-AE achieves the lowest reconstruction error, outperforming the L2 GRU-AE and L3 GRU-AE models. 
Moreover, compared with the L1 AE and L2 AE, the L3 AE attains lower reconstruction error, as the limited semantic compression capability of L1 and L2 AEs leads to instability and divergence in the closed-loop system. 
For state estimation performance, the L2 GRU-AE achieves the best estimation accuracy among all schemes. 
Compared with L1, the improvement of L2 mainly stems from the better match between the semantic encoder and the Kalman filter, as these two components are jointly trained, and the reconstruction noise of the encoded observation $\hat{\bm{y}}$ is explicitly estimated and incorporated during filtering.
The L1 GRU-AE also outperforms the L3 GRU-AE, since the L3 model focuses on optimizing the control objective without explicitly considering the estimation of uncontrollable targets, resulting in higher reconstruction and estimation errors. 
Regarding \ac{lqr} cost, the L3 GRU-AE achieves the lowest overall \ac{lqr} cost among all baseline schemes, only 4.8\% higher than that of the rate-unconstrained \ac{lqr} benchmark. 
Moreover, even when using the L3 AE, its \ac{lqr} cost remains lower than that of the L1 and L2 GRU-AE models. This demonstrates that, compared with the L1 and L2 semantic levels, control-semantic compression at L3 more effectively extracts task-relevant features and substantially improves control efficiency under communication constraints.

\subsection{Joint Evaluation of Compression and Rate Adaptation}
In this subsection, we jointly evaluate the semantic compression and rate adaptation modules to demonstrate the effectiveness of the complete closed-loop \ac{scc} framework and to provide insights into the learned resource allocation strategies across different semantic levels.
We consider a setup with 4 sensors, where sensors~1 and~2 observe only the controllable targets (i.e., the last 4 rows of $\bm{C}^s$ are set to 0), while sensors~3 and~4 observe only the uncontrollable targets (i.e., the first 4 rows of $\bm{C}^s$ are set to 0). The compression dimension of each sensor is set to 1.
To simplify the scenario, the initial velocities are fixed as $v_x^{\text{ctl,init}}=1$, $v_y^{\text{ctl,init}}=0$, $v_x^{\text{unctl,init}}=0$, and $v_y^{\text{unctl,init}}=1$, implying that the desired trajectories are predetermined. 
The simulation spans 500 time slots, divided into four distinct scenarios:
In scenario~1 (slots~1–125), the sensing states of the four sensors are [good/bad/good/bad];
in scenario~2 (slots~126–250), the states are [good/bad/bad/good];
in scenario~3 (slots~251–375), the states are [bad/good/good/bad];
and in scenario~4 (slots~376–500), the states are [bad/good/bad/good].

Fig.~\ref{fig:result_all} compares the system performance under different long-term communication budgets, where the x-axis represents the average bits available across all sensors. We compare the results obtained using equal rate allocation across sensors and time against those achieved by adaptive allocation under different semantic levels. 
Fig.~\ref{fig:result_all}\subref{fig:result_all_figure1} reports the reconstruction error, Fig.~\ref{fig:result_all}\subref{fig:result_all_figure2} presents the state estimation error, and Fig.~\ref{fig:result_all}\subref{fig:result_all_figure3} shows the resulting \ac{lqr} cost.
For the L1 level, when the average communication budget is 4~bits, the equal-allocation scheme leads to divergence of the closed-loop control system, resulting in a large reconstruction error. In contrast, the proposed GRU-AE+PPO approach achieves the lowest reconstruction error among all methods, although its estimation error and \ac{lqr} cost are higher than those of the L2 and L3 schemes. For the L2 level, the proposed approach achieves the best state estimation accuracy across all methods except the UC. For the L3 level, the proposed method achieves the lowest \ac{lqr} cost among all other schemes, except for the UC scheme. With an average long-term communication budget of 4~bits, it yields only a 33.0\% performance loss compared with the UC scheme, while reducing the \ac{lqr} cost by 71.5\% and 14.1\% compared with the L1 and L2 schemes, respectively.
These results demonstrate that in closed-loop \ac{scc} systems, reconstruction error, estimation error, and \ac{lqr} cost are interrelated but exhibit different sensitivities to design objectives.
Therefore, optimizing for specific semantic levels can effectively reshape these trade-offs, underscoring the importance of goal-oriented semantic design in distributed closed-loop systems.

\begin{table}[tbp]
\centering
\caption{Average Bits under Different Semantic Levels and Scenarios}
\label{table:action}
\begin{tabular}{
|>{\centering\arraybackslash}p{1.0cm}
|>{\centering\arraybackslash}p{0.8cm}
|>{\centering\arraybackslash}p{1.1cm}
|>{\centering\arraybackslash}p{1.1cm}
|>{\centering\arraybackslash}p{1.1cm}
|>{\centering\arraybackslash}p{1.1cm}
|}
\hline
\textbf{Semantic Level}
& \textbf{Sensor}
& \textbf{Avg. Bits (Scen.~1 [g/b/g/b])}
& \textbf{Avg. Bits (Scen.~2 [g/b/b/g])}
& \textbf{Avg. Bits (Scen.~3 [b/g/g/b])}
& \textbf{Avg. Bits (Scen.~4 [b/g/b/g])}
\\ \hline
\multirow{4}{*}{L1}
&  1 & 1.34 & 1.31  & 1.36  & 1.38  \\ \cline{2-6}
&  2 & 1.31 &  1.39 & 1.35 & 1.30  \\ \cline{2-6}
&  3 & 0.61 & 0.67 & 0.62  & 0.65  \\ \cline{2-6}
&  4 & 0.72  & 0.67  & 0.71  & 0.70 \\ \hline
\hline

\multirow{4}{*}{\centering L2}
&  1 & 1.57 & 1.57 & 0.52 & 0.55 \\ \cline{2-6}
&  2 & 0.43 & 0.44 & 1.64 & 1.62 \\ \cline{2-6}
&  3 & 1.32 & 0.36 & 1.50 & 0.39 \\ \cline{2-6}
&  4 & 0.64 & 1.54 & 0.55 & 1.47 \\ \hline
\hline

\multirow{4}{*}{\centering L3}
&  1  &  2.31 & 2.26 & 1.34 & 1.39  \\ \cline{2-6}
&  2  &  1.60 & 1.68 & 2.51 & 2.55  \\ \cline{2-6}
&  3  &  0.10 & 0.08 & 0.14 & 0.12 \\ \cline{2-6}
&  4  &  0.06 & 0.05 & 0.08 & 0.02  \\ \hline
\end{tabular}
\end{table}

Table~\ref{table:action} compares the average bit allocation across 4 sensors under different semantic levels and sensing scenarios, with an average communication budget of 4~bits.
At L1, sensors~1 and~2, which observe the controllable targets, are allocated more communication resources than sensors~3 and~4, which observe the uncontrollable targets.
This is because, although the L1 objective targets only observation reconstruction, insufficient transmitted information can cause system divergence, which drives the observations beyond the network’s trained distribution and leads to larger reconstruction errors.
Moreover, sensors observing the same target receive nearly equal rate allocations regardless of their sensing states, indicating that observation noise has little impact on L1 resource allocation in our experimental setup.
At L2, the RL agent tends to allocate more transmission resources to sensors in a \textit{good} state, since these sensors with higher observation quality provide richer semantic information for accurate state estimation.
At L3, the RL agent assigns nearly all available communication resources to sensors~1 and~2, which monitor the controllable targets, while also prioritizing sensors in the \textit{good} state.
This behavior demonstrates that, under the L3 objective, the RL agent effectively allocates communication resources according to the closed-loop control goal (i.e., \ac{lqr} cost), thereby emphasizing task-relevant semantic information. Overall, the rate allocation variance across sensors increases from L1 to L3, reflecting the progressively stronger task-awareness of the proposed framework. 
Specifically, the L3 level shows the most targeted resource allocation, concentrating nearly all bandwidth on controllable targets, whereas L1 distributes rates more uniformly across sensors.

\section{Conclusion}
In this paper, we present a unified framework for goal-oriented semantic communication in closed-loop distributed \ac{scc} systems. 
We define three semantic error levels (L1: observation reconstruction, L2: state estimation, and L3: control) to optimize communication, sensing, and control objectives. 
Based on this, we propose a unified goal-oriented semantic compression and rate adaptation framework that is applicable to different semantic error levels and optimization goals across the \ac{scc} loop. 
We further use a distributed \ac{lqr} system as a case study to illustrate the impact of different semantic levels on overall system performance. 
Within this framework, a GRU-based \ac{ae} and a \ac{ppo}-based rate adaptation algorithm are developed to realize semantic compression and dynamic rate allocation across sensors. 
Experimental results on the rate-limited \ac{lqr} system verify that the proposed hierarchical framework effectively captures task-relevant semantics and achieves efficient closed-loop control. Moreover, the learned rate allocation strategies differ across semantic levels, with L3 showing the strongest task awareness by prioritizing controllable targets and focusing bandwidth on control-relevant information.
Future work includes the extension of this framework to more complex nonlinear systems by incorporating realistic wireless channel effects and multi-agent coordination.

\bibliographystyle{IEEEtran}
\bibliography{IEEEabrv, ref}

\end{document}